\documentclass[aps,prl,notitlepage,superscriptaddress,twocolumn,groupedaddress,showpacs,nofootinbib]{revtex4}

\usepackage{amsmath}

\usepackage{amssymb}
\usepackage{graphicx,color}

\usepackage{amsmath,amsthm}
\usepackage{txfonts}
\usepackage{mathrsfs}

\usepackage{bm}

\newcommand{\Ref}[1]{(\ref{#1})}

\newtheorem{Theorem}{Theorem}[section]
\newtheorem{Definition}{Definition}[section]
\newtheorem{Lemma}[Theorem]{Lemma}

\newcommand{\Z}{\mathbb{Z}}
\newcommand{\R}{\mathbb{R}}
\newcommand{\C}{\mathbb{C}}

\newcommand{\half}{\frac{1}{2}}

\newcommand{\ccirc}{\kern0.2ex\vcenter{\hbox{$\scriptstyle\circ$}}\kern0.2ex}

\newcommand{\rank}{\text{rank}}

\newcommand{\inv}[1]{\mathrm{Inv}_{SU(2)}[#1]}

\newcommand{\Su}{\mathrm{SU}(2)}

\def\be{\begin{eqnarray}}
\def\ee{\end{eqnarray}}

\newcommand{\cc}{\mathcal C}

\newcommand{\cf}{\mathcal F}

\newcommand{\ch}{\mathcal H}

\newcommand{\cj}{\mathcal J}
\newcommand{\ck}{\mathcal K}

\newcommand{\cn}{\mathcal N}

\newcommand{\calr}{\mathcal R}

\newcommand{\cx}{\mathcal X}

\newcommand{\sm}{\mathscr{M}}

\newcommand{\ff}{\mathfrak{f}}

\renewcommand{\a}{\alpha}
\renewcommand{\b}{\beta}
\newcommand{\g}{\gamma}

\renewcommand{\d}{\delta}
\newcommand{\eps}{\varepsilon}

\newcommand{\sig}{\sigma}

\renewcommand{\l}{\lambda}

\renewcommand{\o}{\omega}
\renewcommand{\O}{\Omega}
\renewcommand{\t}{\tau}

\newcommand{\rmd}{\mathrm d}

\newcommand{\lt}{\left}
\newcommand{\rt}{\right}

\newcommand{\lag}{\left\langle}
\newcommand{\rag}{\right\rangle}

\newcommand{\tr}{\mathrm{tr}}

\newcommand{\sgn}{\mathrm{sgn}}
\newcommand{\vth}{\vartheta}

\newcommand{\sn}{\mathscr{N}}

\newcommand{\uab}{U_{ab}}
\newcommand{\ucd}{U_{cd}}

\begin{document}

\sloppy

\title{\bf Emergent 4-dimensional linearized gravity from spin foam model}

\author{Muxin Han}
\affiliation{Department of Physics, Florida Atlantic University, 777 Glades Road, Boca Raton, FL 33431, USA}
\affiliation{Institut f\"ur Quantengravitation, Universit\"at Erlangen-N\"urnberg, Staudtstr. 7/B2, 91058 Erlangen, Germany}

\author{Zichang Huang}
\affiliation{Department of Physics, Florida Atlantic University, 777 Glades Road, Boca Raton, FL 33431, USA}

\author{Antonia Zipfel}
\affiliation{Department of Physics, Florida Atlantic University, 777 Glades Road, Boca Raton, FL 33431, USA}


\begin{abstract}
Spin Foam Models (SFMs) are covariant formulations of Loop Quantum Gravity (LQG) in 4 dimensions. This work studies the perturbations of SFMs on a flat background. It demonstrates for the first time that \emph{smooth} curved spacetime geometries satisfying Einstein equation can emerge from \emph{discrete} SFMs under an appropriate low energy limit, which corresponds to a semiclassical continuum limit of SFMs. In particular, we show that the low energy excitations of SFMs on a flat background give all smooth solutions of linearized Einstein equations (spin-2 gravitons). This indicates that at the linearized level, classical Einstein gravity is indeed the low energy effective theory from SFMs. Thus our result heightens the confidence that covariant LQG is a consistent theory of quantum gravity. As a key technical tool, a regularization/deformation of the SFM is employed in the derivation. The deformation parameter $\delta$ becomes a coupling constant of a higher curvature correction term to Einstein gravity from SFM. 



\end{abstract}


\maketitle

\section{Introduction}

The spin foam program is a covariant approach towards a nonperturbative and background-independent quantum theory of gravity \cite{Reisenberger:1996pu,rovelli2014covariant,Perez2012,KKL}. Spin foam models (SFMs), therefore, provide a powerful formalism to analyze the dynamics of Loop Quantum Gravity (LQG) \cite{book,review,review1,book1}. As state-sum lattice models inspired by topological quantum field theory, SFMs are a LQG analog of Feynman path integral description of quantum gravity \cite{Ponzano1968,Archer:1991rz}. In particular they describe the histories of evolving quantum geometries of space \cite{Reisenberger:1996pu,Kisielowski:2018oiv}. The study of SFMs has uncovered many remarkable properties in the last two decades. Amongst others, SFMs are finite in presence of cosmological constant \cite{QSF,QSF1} and have an interesting semiclassical behavior that relates to General Relativity (GR) \cite{semiclassical,CFsemiclassical,HZ,HZ1,HHKR,propagator,propagator1,Han:2017xwo}. Moreover, SFMs are well-behaved at curvature singularities \cite{Han:2016fgh}. This enables us to study singularities in a concrete quantum gravity model. The above properties make SFMs stand out among lattice quantum gravity models.

The semiclassical consistency is one of the most crucial requirements for a candidate quantum gravity theory. Recent results  show that SFMs give rise to \emph{discrete} spacetime geometries in a \emph{large spin limit} (e.g. \cite{semiclassical,CFsemiclassical,HZ}). The discreteness of the geometries is a consequence of the lattice dependence of SFMs. If SFMs do indeed qualify as models of quantum gravity, then there should also exist a continuum limit under which smooth general relativity arises as an effective low energy theory. The construction of such a limit has been a long standing issue in SFMs  \cite{Dittrich2014,Oriti2007,Bonzom:2015ans,Bahr:2016hwc}.

In this paper, we show for the first time that smooth solutions of 4-dimensional Einstein equation emerge from SFMs under an appropriate semiclassical continuum limit (SCL). The limit combines the large spin limit and lattice refinement in a coherent manner; it also can be interpreted as a low energy limit of SFMs. We focus on the perturbations of SFMs on a flat background, and find the low energy excitations from the SCL give all smooth solutions of linearized Einstein equation. Thus the low energy effective theory of SFMs yields classical Einstein theory at the linearized level.

This work can be also understood along the lines of the emergent gravity program. An idea in this program is that gravity, which is geometrical and smooth, might emerge as the low energy excitations from fundamentally entangled qubits (or generally qudits), which are algebraic and discrete \cite{2011JSP...145..891E,Swingle:2009bg,Pastawski:2015qua,Qi1,Wen:2018dqi,Qi:2018shh}. In this paper, we show that SFMs can be rewritten in terms of spacetime tensor networks (TNs), whose fundamental degrees of freedom (DOFs) are entangled qudits at different spacetime locations. Therefore, our results prove to be a working example for the above idea.

\section{Spin foam models:} \label{secSFMs}

\begin{figure}
\begin{center}
\includegraphics[width = 0.38\textwidth]{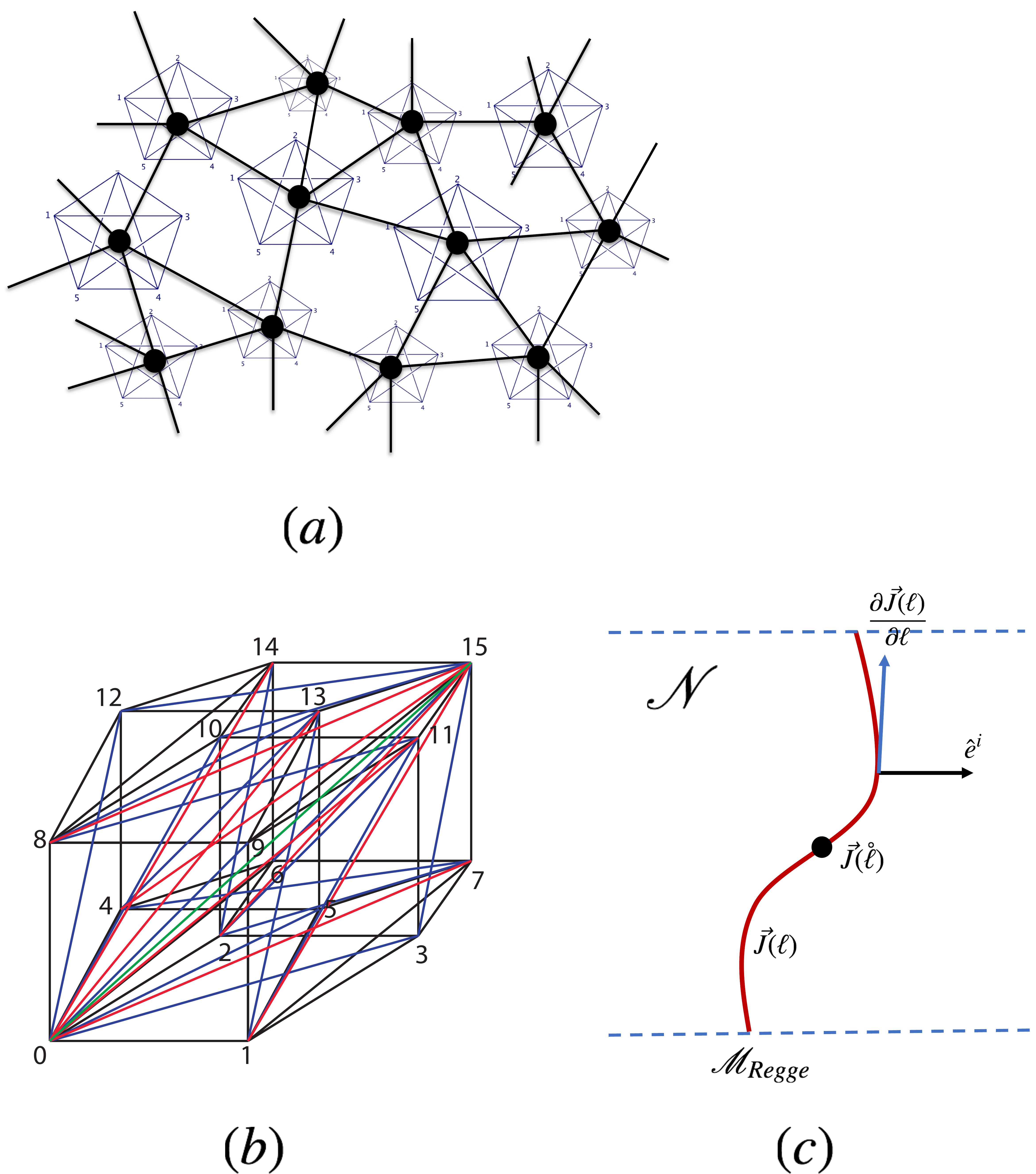}
\end{center}
\caption{(a) The 5-valent vertex in a 4-simplex illustrates a rank-5 tensor $|A_\sig\rangle$. Gluing 4-simplices $\sig$ in $\ck$ gives a tensor network $\mathrm{TN}(\ck,\vec{J})$, where each link associates to a maximally entangled state of a pair of $i_\t$s. (b) A triangulation of the hypercube. The 4d hypercubic lattice with the triangulated hypercube makes $\ck$. (c) An illustration of the neighborhood $\sn$ (the region bounded by blue dashed lines) in the space of $\vec{J}$. The red curve illustrates $\sm_{Regge}$, including $\vec{J}(\ell)$ as the perturbation of $\vec{J}(\mathring{\ell})$. The black and blue arrows are basis vectors $\hat{e}^i(\ell)$ and $\partial \vec{J}(\ell)/\partial\ell$, transverse and tangent to $\sm_{Regge}$.}
\label{cube}
\end{figure}

SFMs are defined over 4-dimensional (4d) simplicial triangulations $\ck$, which are obtained by gluing 4-simplices $\sig$ along their common tetrahedra $\t$ quite similar to the gluing of tetrahedra in 3d triangulation or triangles in 2d triangulation. Thus a triangulation $\ck$ consists of simplices $\sig$, tetrahedra $\t$ (boundaries of $\sig$s), triangles $f$ (boundaries of $\t$s), edges (boundaries of $f$s) and vertices. Our analysis focuses on $\ck$ adapted to a hypercubic lattice in $\R^4$ in such a way that each hypercube is triangulated identically by 24 4-simplices (see FIG.\ref{cube}(b)). The same triangulation has been employed in e.g. \cite{0264-9381-5-12-007,Rocek:1982tj} to study perturbations on a flat background. Here $\ck$ is a finite lattice with boundary in a region of $\R^4$.

A SFM is obtained by associating a state sum, 
\be
Z(\ck)=\sum_{\vec{J},\vec{i}}\prod_fA_f(J_f)\prod_\sig A_\sig(J_f,i_\t),
\ee 
to $\ck$ and can be interpreted as the path integral of a triangulated manifold (here $\R^4$). In the above state sum, each triangle $f$ is colored by an SU(2) representation $J_f\in \Z_+/2$ and each tetrahedron $\t$ is colored by an SU(2) intertwiner (invariant tensor) $i_\t$. They are quantum numbers labelling histories of LQG quantum geometry states, which are the intermediate states of the path integral. $J_f,i_\t$ can be related to the area of $f$ and the shape of $\t$ in the semiclassical interpretation \cite{Rovelli1995,ALarea,shape}. The dynamics of the model is captured in the \emph{4-simplex amplitudes} $A_\sig(J_f,i_\t)\in\C$ associated to each $\sig$. In particular,  $A_\sig(J_f,i_\t)\in\C$ describes the local transition between the quantum geometry states labelled by $\{J_f,i_\t\}$ for $f,\t$ on the boundary of $\sig$. The weights of the spin sum $A_f(J_f)$ is the face amplitudes. 

The amplitudes $A_\sig(J_f,i_\t)$ depend linearly on the intertwiners $i_\t$ and thus are rank-5 tensors on intertwiner spaces. The 4-simplices in $\ck$ are glued by identifying a pair of $\t$s in $\sig$ and $\sig'$. This implies that $\sum_{\vec{i}}A_\sig$ is equivalent to the inner products between the tensors $|A_\sig\rangle$ at all $\sig$s and the maximally entangled states $|\t\rangle=\sum_{i_\t}|i_\t\rangle\otimes|i_\t\rangle$, where $i_\t$ are shared by pairs of $\sig$s. This yields a spacetime tensor network (TN) (FIG.\ref{cube}(a))
\be
\mathrm{TN}(\ck,\vec{J}):=\otimes_\t\langle\t| \otimes_\sig|A_\sig(J_f)\rangle.
\ee
Note that the entangled intertwiners (the qudits) are the fundamental DOFs of the TN. Moreover the state sum $Z(\ck)$ can now be expressed in terms of these TN, that is, $Z(\ck)=\sum_{\vec{J}}\mathrm{TN}(\ck,\vec{J})\prod_fA_f(J_f)$. More details on SFM and TN are given in Appendix A.

The following demonstrates that smooth Einstein solutions can emerge from the fundamentally entangled intertwiners. Thus it realizes the idea of emergent gravity from entangled qubits. In order to show this, we employ the integral representation of $Z(\ck)$ \cite{semiclassicalEu,CFsemiclassical,HZ1}:
\be
Z(\ck)=\sum_{\vec{J}}\prod_fA_f(J_f)\int\lt[\rmd X\rt]\,e^{\sum_{f}J_f F_f\lt[X\rt]}.\label{Z00}
\ee
Here $F_f$ is a function that only depends on a set of spinfoam variables 
\be
X\equiv(g_{\sig\t}^\pm,\xi_{\t f}),\label{X}
\ee
which includes $(g^+_{\sig\t},g^{-}_{\sig\t})\in\,$Spin(4) at pairs of $(\sig,\t)$ with $\t\subset\sig$, and $\xi_{ef}\in \mathbb{CP}^1$ at pairs of $(\t,f)$ with $f\subset\t$. The details of $A_f$ and $F_f$ depend on the specific SFM. $A_f$ is often chosen as $(2J_f+1)^{\a_f}$. Here for the purpose of large-$J$ analysis, we set $A_f$ as $(2J_f)^{\a_f}$.

Here, we focus on the Euclidean Engle-Pereira-Rovelli-Livine/Freidel-Krasnov (EPRL/FK) model ($\gamma < 1$) \cite{EPRL,FK} where
\be
F_f&=&\sum_{\sig, f\subset\sig}\Big[(1-\g)\ln\lag{\xi_{\t f}}\big|(g^-_{\sig\t})^{-1}g^-_{\sig\t^\prime}\big|{\xi_{\t^\prime f}}\rag\nonumber\\
  &&\quad+(1+\g) \ln\lag{\xi_{\t f}}\big|(g^+_{\sig\t})^{-1}g^+_{\sig\t'}\big|{\xi_{\t' f}}\rag\Big],
  \ee
but our results can be generalized to other SFMs, e.g., \cite{Engle2013,Kaminski:2017eew,HHKR}.

$\g=p/q$ ($p,q\in\Z_+$). $J\in q\Z$ when $p+q$ is odd $J\in q\Z$), and $J\in q\Z/2$ when $p+q$ is even. We assume even $p+q$ in the following computation. One may replace $q\to 2q$ to obtain results for odd $p+q$.

The integrand in Eq.\Ref{Z00} is manifest periodic by $F_f\sim F_f+4\pi i/q$. So we set $\mathrm{Im}(F_f)\in[-2\pi/q,2\pi/q]$. It also has a discrete symmetry $g^+_{\sig\t}\to\pm g^+_{\sig\t}$ and independently $g^-_{\sig\t}\to\pm g^-_{\sig\t}$. The transformation simultaneously shift $F_f\to F_f+\pi i(1\pm\g)$ for all $f\subset\t$, and gives a factor $e^{2\pi i\sum_{f\subset\t}J_f^\pm}$ ($J^\pm_f=\frac{1\pm\g}{2}J_f\in\Z/2$). This factor equals $1$ because $\sum_{\vec{J}}$ is constrainted by $\sum_{f\subset\t}J_f^\pm\in\Z$:
\be
\sum_{\vec{J}}&=&\prod_{f,\t,\pm}\sum_{J_f\in\mathbb{N}/2}\sum_{n^\pm_\t\in\Z}\delta_{n^\pm_\t,\sum_{f\subset\t}J_f^\pm}\nonumber\\
&=&\frac{1}{2^{N_\t}}\prod_{f,\t,\pm}\sum_{J_f\in\mathbb{N}/2}\sum_{m^\pm_\t\in{0,1}}e^{2\pi i m_\t^\pm\sum_{f\subset\t}J_f^\pm }.\label{jsum}
\ee
where $N_\t$ is the number of $\t\subset\ck$. Although the integral vanishes for $\vec{J}$ violating this constraint, it is useful to explicitly impose this constraint to $\sum_{\vec{J}}$ for the purpose of asymptotic analysis of the integral.

\section{spin sum and Regularization:} 

LQG predicts that the geometrical areas are fundamentally discrete at the Planck scale. The area spectrum  \cite{Rovelli1995,ALarea} relates to the spins via $\mathbf{a}_f=\g\sqrt{J_f(J_f+1)}\ell_P^2$, where $\g\in\R$ is the Barbero-Immirzi parameter and $\ell_P^2\equiv8\pi G_N\hbar$. 
Since the semiclassical area $\mathbf{a}_f\gg\ell_P^2$ implies $J_f\gg1$, the semiclassical analysis of SFMs is build on uniformly large (but finite) spins $J_f=\l j_f$ where $\l\gg1$ is the typical value of the spins.

For the following argument, it is important to note that small perturbations $\mathring{J}+\delta J$ of a given background spins $\mathring{J}\sim\l\gg1$ will still be inside this large-$J$ regime. Moreover, the sum $\sum_{\vec{J}}$ can be replaced through an integral by Poisson resummation formula. by Eq.\Ref{jsum},
\begin{widetext} 
\be
Z(\ck)=\sum_{k_f,m_\t^\pm\in\Z/\sim}\!\!\frac{2^{N_f-N_\t}}{q^{N_f}}\!\!\int_0^\infty\lt[\rmd J\rt]\int\lt[\rmd X\rt]\prod_fA_f(J_f)\,e^{\sum_{f}J_f \lt(F_f\lt[X\rt]+4\pi i \frac{k_f}{q}+\pi i \sum_{\pm}m_\t^\pm \frac{q\pm p}{q}\rt)},\label{Z0}
\ee
\end{widetext}
where $N_f$ denotes the number of internal $f$s in $\ck$. $4\pi i \frac{k_f}{q}+2\pi i \sum_{\pm}m_\t^\pm \frac{q\pm p}{q}$ in the exponent and $\sum_{{k}_f,m_\t}$ manifest the periodicity of the integrand discussed above. $\sum_{k_f,m_\t\in\Z/\sim}$ sums $k_f,m_f\in\Z$ modulo an equivalence because the exponent has gauge transformations
\be
\lt\{k_f\rt\}_{f\subset\t}\to\lt\{k_f+\sum_\pm({q\pm p})\frac{M^\pm_\t}{2}\rt\}_{f\subset\t},\quad m_\t^\pm\to m^\pm_\t-2M_\t^\pm,\label{redund}
\ee
where $M_\t\in\Z$ and $q\pm p$ are even numbers. Note that in Eq.\Ref{Z0} we only focus on the terms from internal $f$ and neglect the boundary terms, since they do not involve in most of the following analysis.

Eq.\Ref{Z0} treats $J$s as continuous variables. From previous results e.g. \cite{semiclassicalEu,CFsemiclassical,HZ1} follows that there is a subspace of large $\vec{J}\in \R^{N_f}$ that determine classical triangle areas. These spin conigurations are called \emph{Regge-like} and satisfy the triangle area-length relation 
\be
\g {J}_f(\ell)=\frac{1}{4}\sqrt{2(\ell^2_{ij}\ell^2_{jk}+\ell^2_{ik}\ell^2_{jk}+\ell^2_{ij}\ell^2_{ik})-\ell_{ij}^4-\ell_{ik}^4-\ell_{jk}^4}.\label{area-length}
\ee  
The right-hand side determines the area ${\mathbf{a}_f(\ell)}/{\ell_P^2}$ of the triangle $f$ in terms of $\ell_{ij},\ell_{ik},\ell_{jk}$ being the lengths (in $\ell_P$ unit) of 3 edges of a triangle. Since there are less edges than triangles in the bulk of $\ck$, Regge-like spins form a proper subset and Eq.\Ref{area-length} defines an embedding map $\R^{N_\ell}\hookrightarrow\R^{N_f}$. $N_\ell$ is the number of internal edges in $\ck$.


Here, we want to consider perturbations on a flat (triangulated) hypercubic lattice with constant spacing $(\g\l)^{1/2}$ (in $\ell_P$ unit), which fixes all edge lengths $\mathring{\ell}$ in $\ck$, e.g. $(\g\l)^{-1/2}\mathring{\ell}=1,\sqrt{2},\sqrt{3},2$ for the cube edges, face diagonals, body diagonals, and hyperbody diagonals. These edge-lengths in turn determine the Regge-like spins $\mathring{\vec{J}}=\vec{J}(\mathring{\ell})$ by Eq.\Ref{area-length}. The flat triangulated hypercubic lattice geometry is a large-$J$ critical point of the SFM and determines the critical data $\mathring{X}$. In this paper, we focus on the perturbations
\be
(J,X)=(\mathring{J}+\delta J,\mathring{X}+\delta X).
\ee 
When $\mathring{J}\sim\l\gg1$, $\mathring{J}+\delta J$ are also large. The perturbations restricts us in the large-$J$ regime of $Z(\ck)$. By considering perturbations of the flat geometry, we would like to extract solutions of equations of motion from SFM, and find their continuum limit to give the smooth linearized Einstein gravity. 

For the study of perturbations around $\mathring{\vec{J}}$, it is sufficient to consider a neighborhood $\sn\subset\R^{N_f}$ of $\mathring{\vec{J}}$. $\sn$ is constructed as follows: Firstly, smooth perturbations $\ell=\mathring{\ell}+\delta\ell$ and the embedding Eq.\Ref{area-length} define a submanifold $\sm_{Regge}\subset\R^{N_f}$ of dimension $N_\ell$. We choose $\hat{e}^i$ ($i=1,\cdots,N_f-N_\ell$) basis vectors transverse to $\sm_{Regge}$. All 
\be
\vec{J}=\vec{J}(\ell)+\sum_{i=1} t_{i}\hat{e}^i
\ee 
defines $\sn$, with $\vec{J}(\ell)\in \sm_{Regge}$ and $t_i\in \R$. $(\ell,t^i)$ form a local coordinate system in $\sn$ (see FIG.\ref{cube}(c)). $J$s with ${t}^i\neq0$ is called \emph{non-Regge-like}. $\hat{e}^i$ can be chosen as constant vectors transverse to $\sm_{Regge}$ since we focus on a neighborhood at $\mathring{\ell}\in\sm_{Regge}$ (the space of $\vec{J}$ is a flat space $\R^{N_f}$). For instance, we can choose $\hat{e}^i$ to be vectors normal to $\sm_{Regge}$ at $\mathring{\ell}$, and extend every $\hat{e}^i$ to a constant vector field trivially by parallel transport in $\R^{N_f}$. $\hat{e}^i$ are transverse to $\sm_{Regge}$ in a neighborhood of $\mathring{\ell}$.

The integral over $\vec{J}$ can now be split into transverse and Regge-like part as well. That is, $\int\rmd\vec{J}=\int[\rmd\ell\rmd t]\,\cj(\ell) $, where the Jacobian $\cj(\ell)=|\partial \vec{J}(\ell)/\partial\ell,\hat{e}^i|$. $\cj(\ell)$ only depends on $\ell$ because $\hat{e}^i$ are constant vectors. We regularize the transverse integral $\int_{-\infty}^{\infty} t_i$ by inserting a Gaussian factor parametrized by $0<\delta\ll1$:  
\be
\label{mod}
\int\rmd\vec{J}=\int[\rmd\ell\rmd t]\,\cj(\ell)\to \int[\rmd\ell]\,\cj(\ell)\int[\rmd t]\,e^{-\frac{\delta}{4}\sum_{i}t_i^2}.
\ee
The $t$-integral has a lower bound since $J\in[0,\infty)$. But extending the $t$-integral to $-\infty$ only add a negligible contribution when $\mathring{J}$ is large.

Inserting Eq.\Ref{mod} to $Z(\ck)$ defines
\begin{widetext}
\be
Z_\delta(\ck):=\frac{2^{N_f-N_t}}{q^{N_f}}\sum_{k_f,m_\t^\pm\in\Z/\sim}\int_0^\infty[\rmd\ell]\,\cj(\ell)\int[\rmd X\rmd t]\,e^{-\frac{\delta}{4}\sum_{i}t_i^2}\prod_fA_f(J_f)\,e^{\sum_{f}J_f \lt(F_f\lt[X\rt]+4\pi i \frac{k_f}{q}+\pi i \sum_{\pm}m_\t^\pm \frac{q\pm p}{q}\rt)},\label{Z01}
\ee
\end{widetext}
where we can interchange $\int[\rmd X]$ and $\int[\rmd t]$ since $\int[\rmd X]$ is over a compact space and $e^{-\frac{\delta}{4}\sum_{i}t_i^2}$ decays at infinity. 

The regulator $\delta$ plays a key role in our work. The following explains several roles played by this regularization:

\begin{itemize}

\item Inserting the Gaussian modifies the sum over spins $\sum_{J_f=0}^\infty$ along certain direction in the space of spins. Indeed if we perform the Poisson resummation backward after inserting the Gaussian in Eq.\Ref{Z0}, using the relation $\sum_{k\in\Z}e^{2\pi i k x}=\sum_{n\in\Z}\delta(x-n)$ recovering $J\in\Z/2$ and $\sum_{f\subset\t} J^\pm\in \Z$ from continuous $J$ in Eq.\Ref{Z0},
\begin{widetext}
\be
Z_\delta(\ck)&=& \frac{2^{N_f-N_\t}}{q^{N_f}} \int[\rmd\ell]\,\cj(\ell)\int[\rmd X\rmd t]\,e^{-\frac{\delta}{4}\sum_{i}t_i^2}\prod_fA_f(J_f)\,e^{\sum_{f}J_f F_f\lt[X\rt]}\nonumber\\
&\times &\prod_f\sum_{{n}_f\in\Z}\delta\lt({2{J}_f}/{q}-{n}_f\rt)\prod_{\t, \pm}\sum_{n_\t^\pm\in\Z}\delta_{n_\t^\pm,\sum_{f\subset\t} J^\pm_f}\label{Z03}
\ee
\end{widetext}
It is clear that $Z_\delta(\ck)$ modifies $Z(\ck)$ by damping down spins with large $t$ (far away from $\sm_{Regge}$ transversely). Integrating delta functions in Eq.\Ref{Z03} then sending $\delta\to 0$ reduces $Z_\delta(\ck)$ to $Z(\ck)$ (comparing to Eq.\Ref{jsum}). The Gaussian with small $\delta$ is a ``smooth cut-off'' of large spins (in the direction transverse to $\sm_{Regge}$). 

\item $Z_\delta(\ck)$ is a 1-parameter deformation from $Z(\ck)$, and $\delta$ is a parameter deciding how many non-Regge-like $J$s are contributing $Z_\delta(\ck)$. From $\delta=0$ to $\delta\to\infty$, $Z_\delta(\ck)$ has less and less non-Regge-like contribution. $\delta\to\infty$ removes all non-Regge-like contribution from $Z(\ck)$. However in the following discussion, we focus on small $\delta$ and send $\delta\to 0$ in the end. Namely we only ignore spins very far away from $\sm_{Regge}$ and $\mathring{J}$ (at discrete level), which is qualified because we study perturbative effects around $\mathring{J}$, as will be clarified in a moment.

\item Although $Z_\delta(\ck)\neq Z(\ck)$ at the discrete level, they may have the same continuum limit if we turn off the regulator $\delta\to0$ together with refining the lattice $\ck$, as we will do in the following. Eventually the theory of spinfoams should be defined in the continuum limit to remove the triangulation dependence. All physical quantities computed in the continuum limit will not depend on $\delta$. 

\item As we see in a moment, another important role played by $\delta\ll1$ is to make deficit angles $\eps_f$ of emergent Regge geometries to be small but nonzero, as a resolution of the ``flatness problem'' in SFM \cite{frankflat,Perini:2012nd,flatness,Han:2017xwo}. The detailed discussion is given momentarily below Eq.\Ref{small}.

\item It is discussed shortly below that $\delta$ reveals the high curvature corrections to Einstein gravity derived from SFM. It is demonstrated in Eq.\Ref{Z1} and explained shortly below. It is closely related to \cite{Han:2013tap}. $\delta\in[0,\infty)$ is essentially a parameter interpolating from SFMs to quantum Regge calculus.

\item Note that inserting the Gaussian with $\delta$ may not make the SFM finite, since the domain of $\int_0^\infty[\rmd\ell]$ still contains the orbits of the vertex translation group (zero modes in a Hessian matrix in Section \ref{EOLT}). This orbit is non-compact and not restricted by $\delta$.

\end{itemize}

\section{higher curvature correction}

We compute the term in $Z_\delta(\ck)$ at $k_f=m_\t=0$ (all other terms can be obtained by shifting $F_f\to F_f+4\pi i \frac{k_f}{q}+\pi i \sum_{\pm}m_\t^\pm \frac{q\pm p}{q}$):
\be
&&\int\lt[\rmd \ell\rmd X\rt]\,\cj(\ell)\int[\rmd t]\,e^{-\frac{\delta}{4}\sum_{i}t_i^2}\nonumber\\
&\times&\prod_fA_f\lt(J_f(\ell)+\sum_{i=1} t_{i}\hat{e}_f^i\rt)\,e^{\sum_{f}\lt[{J}_f(\ell)+\sum_{i=1} t_{i}\hat{e}_f^i \rt]F_f\lt[X\rt]}
\ee
The $t_i$-integral in $Z(\ck)$ is a Gaussian integral and yields 
\be
\int[\rmd\ell\rmd X]\, e^{\l\lag \vec{j}(\ell),\vec{F}[X]\rag}\,D_\delta(\ell,X),\quad D_\delta= e^{\sum_{i=1}^M\frac{1}{\delta}\langle \hat{e}^i,\vec{F}[X]\rangle^2}\cj'\label{Z1}
\ee
where $\cj'=(\frac{4\pi}{\delta})^M\cj(\ell)\prod_fA_f({J}_f(\ell)+\frac{2}{\delta}\sum_i \hat{e}^i_f\langle \hat{e}^i,\vec{F}[X]\rangle)$. Here, $\vec{F}=\{F_f\}_f$ is treated as a complex $N_f$-dimensional vector, and $\lag\cdot,\cdot\rag$ denotes the Euclidean inner product. Furthermore, we have ignored the boundary terms in the exponent because they are unimportant in the main discussion. 

We can combine the exponent of $D_\delta$ and define an effective action $S_{eff}$ write Eq.\Ref{Z1} as
\be
&&\int[\rmd\ell\rmd X]\, e^{S_{eff}[\ell,X]}\cj'[\ell,X],\quad \text{where}\nonumber\\
&&S_{eff}=\l\lt[\lag \vec{j}(\ell),\vec{F}[X]\rag+\frac{1}{\l\delta}\sum_{i=1}^M\langle \hat{e}^i,\vec{F}[X]\rangle^2\rt].\label{F2}
\ee
where we have written $J_f=\l j_f$ ($\l\gg 1$ is the typical value of $\mathring{J}$ over $\ck$).

In the SFM large-$J$ asymptotics, $F_f=i\g \eps_f$ at a subclass of geometrical large-$J$ critical points of the integral (see e.g \cite{HZ1} and Sections \ref{face} and \ref{angles}), where $\eps_f$ is the deficit angle in Regge geometry. Therefore in $S_{eff}$, $(\delta\l)^{-1}$ is a coupling constant for a $\eps^2_f$ correction, while the first term in $S_{eff}$ reduces to the Regge action at the critical point. The details of this argument is given in the following sections. $\eps_f^2$ term corresponds to higher curvature corrections in Regge calculus \cite{Hamber:1984kx}, although here the $\eps^2_f$ term is likely nonlocal due to the appearance of $\hat{e}^i$.

The above argument is obvious perturbative, because treating the second term in Eq.\Ref{F2} nonperturbatively modifies the critical equation and critical points in the large-$J$ asymptotics. This term contains corrections from SFM degrees of freedom $X$ other than $\eps_f^2$. However in this paper, we still treat this term perturbatively, i.e. we consider the regime 
\be
\l\gg\delta^{-1}\gg1.\label{regime}
\ee
which makes the coupling constant $(\l\delta)^{-1}$ small. All non-$\eps_f^2$-corrections are restored in the perturbative expansion in the coupling constant. The nonperturbative study of $S_{eff}$ beyond the above regime will be reported in the future. Note that given any arbitrarily small $\delta\neq 0$, the above regime always exists because $J_f$ is summed toward infinity.

Recall that $\frac{1}{\delta}\sum_{i=1}^M\langle \hat{e}^i,\vec{F}[X]\rangle^2$ in $S_{eff}$ comes from the $t$-integral which are contributions from non-Regge-like $J$s. Treating this correction term perturbatively in $S_{eff}$ means that we treat the contribution from non-Regg-like $J$s perturbatively. It reflects our proposal mentioned above that we focus on perturbations at $\mathring{J}$.

$Z_\delta(\ck)$ has the following gauge symmetry:
\begin{itemize}

\item \emph{Continuous}: The following transformation leave all $F_f(X)$ invariant: (1) A diagonal Spin(4) action at $\sig$, $g_{\sig\t}^\pm\to h^\pm_\sig g_{\sig\t}^\pm $ for all $\t\subset\sig$ by $(h^+_\sig,h^-_\sig)$; (2) At an internal $\t$, $|\xi_{\t f}\rangle\to h_{\t}|\xi_{\t f}\rangle$ and $g^\pm_{\sig \t}\to g^\pm_{\sig \t}h_{\t}^{-1}$ for all $\sig$ having $\t$ at boundaries; and (3) $|\xi_{\t f}\rangle\to e^{i\theta_{\t f}}|\xi_{\t f}\rangle$ at any internal $|\xi_{\t f}\rangle$.

\item \emph{Discrete}: $g^+_{\sig\t}\to\pm g^+_{\sig\t}$ and independently $g^-_{\sig\t}\to\pm g^-_{\sig\t}$ shift $F_f\to F_f+\pi i(1\pm\g)$ for all $f\subset\t$. Simultaneously $m_\t^\pm\to m_\t^\pm -1$ leaves the integrand invariant. 

\end{itemize}

\section{Critical Equations and Geometrical Correspondence}

Since the exponent in Eq.\Ref{Z1} scales linearly in $\l$, we can apply the stationary phase method to Eq.\Ref{Z1}. As long as the exponent in $D_\delta$ is subleading, we can directly take over the result in \cite{semiclassicalEu,HZ1,CFsemiclassical}.
In the regime of Eq.\Ref{regime}, the dominant contributions of Eq.\Ref{Z1} come from the critical points $(\ell_c,X_c)$, i.e. the solutions of the critical equations $\mathrm{Re}S=\delta_X S=\delta_\ell S=0$, of $S=\langle \vec{j},\vec{F}\rangle$. 

We firstly discuss the subset of critical equations $\mathrm{Re}S=\delta_X S=0$, and postpone discussion of the other critical equation $\delta_\ell S=0$ to Section \ref{EOMASDA}. $\mathrm{Re}S=\delta_X S=0$ have been studied extensively in the spinfoam asymptotic analysis on simplicial complexes with fixed $J_f$ e.g. \cite{CFsemiclassical,HZ,HZ1,hanPI}. Subsections \ref{COST} reviews some key results useful in our derivation, while some details is provided in Subsections \ref{face} and \ref{angles}.

\subsection{Classification of solutions to $\mathrm{Re}S=\delta_X S=0$}\label{COST}

Recall Eq.\Ref{X}, $\delta_X S=0$ includes $\delta_{g_{\sig\t}^\pm}S=\delta_{\xi_{\t f}}S=0$. But $\delta_{\xi_{\t f}}S=0$ is implied by $\mathrm{Re}S=0$ so doesn't give a new constraint \cite{HZ1}. These critical equations are equivalent to the following equations:
\be
\hat{g}_{\sig\t}^\pm\hat{n}_{\t f}=\hat{g}_{\sig\t'}^\pm\hat{n}_{\t' f},\quad \sum_{f\subset\t} j_f\eps_{\t f}(\sig)\hat{n}_{\t f}=0\label{critical}
\ee
where $\hat{g}_{\sig\t}^\pm\in \mathrm{SO(3)}$ is the 3-dimensional representation of ${g}_{\sig\t}^\pm$, and $\hat{n}_{\t f}=\langle\xi_{\t f}|\vec{\sig}|\xi_{\t f}\rangle$ is a unit 3-vector ($\vec{\sig}$ are Pauli matrices). $\eps_{\t f}=\pm1$ satisfying $\eps_{\t f}(\sig)=-\eps_{\t' f}(\sig)$ ($\t\cap\t'=f$) and $\eps_{\t f}(\sig)=-\eps_{\t f}(\sig')$ ($\sig\cap\sig'=\t$). We denote a solution to above equations by
\be
X_c\equiv (g_{\sig \t}^\pm,\xi_{\t f})_c\quad \text{modulo gauge}
\ee
Note that $j_f$ enters as a parameter in these critical equations. A bad choice of $j_f$ may not lead to any solution. But in our case Eq.\Ref{Z1}, $j_f=j_f(\ell)$ implies solutions always exist.   

At $j_f=j_f(\ell)$, there is a subclass $\mathscr{G}$ of solutions $X_c$ which can be interpreted as nondegenerate simplicial geometries on $\ck$. A useful quantity classifying solutions is $\sgn(V_\sig)$ where $V_\sig$ is the oriented 4-volume
\be
\sgn(V_\sig)=\sgn \lt[\det\lt(N_1(\sig),N_2(\sig),N_3(\sig),N_4(\sig)\rt)\rt],
\ee
where $N_{\t}(\sig)$ is the 4d normal of tetrahedron $\t\subset\sig$ outward pointing from $\sig$. $N_{\t}(\sig)$ is computed by $N^0_\t(\sig)\mathbf{1}+iN^i_\t(\sig)\sig_i=g_{\sig\t}^{-}(g_{\sig\t}^{+})^{-1}$ ($\sig_i$ are Pauli matrices). The subclass $\mathscr{G}$ is defined as solutions with $\sgn(V_\sig)\neq 0$.

Eq.\Ref{critical} obviously has a $\Z_2$ symmetry: $\hat{n}_{\t f}\to -\hat{n}_{\t f}$ ($\xi_{\t f}\to J\xi_{\t f}$) globally on the entire $\ck$. $\ck$ triangulates a region in $\R^4$ and has boundary, this symmetry is broken by the boundary condition which fixes $\xi_{\t f}$ at boundary.

The following 1-to-1 correspondence is valid within the subclass $\mathscr{G}$ (see Subsection \ref{face} for a proof, and see \cite{hanPI} for a proof in Lorentzian signature):
\be
&\text{Solutions }X_c\in \mathscr{G}&\nonumber\\
&\updownarrow&\nonumber\\
&\text{4d nondegenerate simplicial geometry on $\ck$}&\nonumber\\
&  \text{and 4-simplex orientations}& \label{equiv0}
\ee  
$X_c$s reconstruct a nondegenerate simplicial geometries on $\ck$ made by geometrical 4-simplices at all $\sig$, while every pair of 4-simplices are glued by sharing a geometrical tetrahedron. Simplicial geometries are parametrized by edge lengths. Some solutions give precisely the simplicial geometry $\ell$ in $j(\ell)$, although some other solutions may give different geometries. But all geometries have the same set of areas $\mathbf{a}_f=\g \l j_f(\ell)\ell_P^2$. $\xi_{\t f}$ in the solution data give tetrahedron face normals $\hat{n}_{\t f}$ of the simplicial geometry.

A simple way to see the appearance of 4-simplex orientations in the above equivalence is that the geometrical data (edge lengths) are invariant under local orthogonal O(4) transformations in $\sig$. Discrete O(4) transformations (parity transformations) acting on the geometry can leads to different $X_c$'s since $X_c$ is only Spin(4) invariant.  

The local parity $\mathbf{P}$ in O(4) leads to the ``cosine problem'' in SFM \cite{semiclassicalEu}. Any $X_c\in\mathscr{G}$ gives $(g_{\sig\t}^+,g_{\sig\t}^-)$ in every $\sig$ with $g_{\sig\t}^+\neq g_{\sig\t}^-$. A party transformation at a $\sig$ flips $g_{\sig\t}^+,g_{\sig\t}^-$ and leaves $\xi_{\t f}$ invariant:
\be
\mathbf{P}_\sig: (g_{\sig\t}^+,g_{\sig\t}^-)\to (g_{\sig\t}^-,g_{\sig\t}^+)
\ee
and maps $X_c$ to another solution $\tilde{X}_c\in\mathscr{G}$ corresponding to the same simplicial geometry. $X_c,\tilde{X}_c$ give opposite 4-orientations to 4-simplex $\sig$, since $\mathbf{P}$ flips the 4-orientation. Local parities gives all orientations in Eq.\Ref{equiv0} on $\ck$. $\sgn(V_\sig)$ characterizes the 4d orientation. $\mathbf{P}_\sig$ gives the parity refection of $N_\t(\sig)$, thus flips $\sgn(V_\sig)$. $\sgn(V_\sig)$ in general not equals to $\sgn(V_\sig')$ for $\sig\neq\sig'$.

The above shows that a solution with $(g_{\sig\t}^+,g_{\sig\t}^-)$ and $\xi_{\t f}$ associates another solution $(g_{\sig\t}^-,g_{\sig\t}^+)$ with the same $\xi_{\t f}$ for the same nondegenerate simplicial geometry. But it is easy to see that $(g_{\sig\t}^+,g_{\sig\t}^+)$ and $(g_{\sig\t}^-,g_{\sig\t}^-)$ with the same $\xi_{\t f}$ are also solutions of Eq.\Ref{critical}. $(g_{\sig\t}^+,g_{\sig\t}^+)$ and $(g_{\sig\t}^-,g_{\sig\t}^-)$ have $\sgn(V_\sig)=0$ so do not belong to the subclass $\mathscr{G}$, and are called BF-type solutions, since they also appears in the asymptotics of SU(2) BF theory.

Another subclass of solution are called vector geometries, which happens when Eq.\Ref{critical} has only a single solution with $(g_{\sig\t},g_{\sig\t})$ in a $\sig$ with some $\xi_{\t f}$. The vector geometry corresponds to a degenerate 4-simplex, and has $\sgn(V_\sig)=0$. Generally speaking, critical equations with $j(\ell)$ ($\ell$ is a nondegenerate simplicial geometry) may still have vector geometry solutions.

The subclass $\mathscr{G}$ of geometrical solutions satisfying Eq.\Ref{equiv0}, BF-type solutions, and vector geometry solutions completely classifies all solutions to Eq.\Ref{critical} on $\ck$ \cite{HZ1,semiclassicalEu}, assuming $\xi_{\t f}$ (internal and at boundary) do not give degenerate tetrahedra. Solutions to Eq.\Ref{critical} with degenerate tetrahedra have not been studied in the literature. Given a generic solution, $\ck$ may need to be divided into regions, such that the solution data restricted into every region are of a single type \cite{HZ,HZ1}.

\subsection{Geometrical Correspondence of Critical Solutions}\label{face}

The following presents a proof of Eq.\Ref{equiv0} of the geometrical correspondence of critical solutions. In this subsection and subsection \ref{angles}, we assume $\ck$ is a generic simplicial complex with or without boundary. The discussions are valid for the triangulation in FIG.\ref{cube}(b), and are also valid for arbitrary triangulations. \\

\noindent
\emph{1. Reconstructing individual 4-simplices:} Given a solution $(g_{\sig\t}^\pm,\xi_{\t f})_c$ (modulo gauge) to Eq.\Ref{critical} with $\vec{J}(\ell)$. We firstly construct five 4-vectors $N_\t(\sig)$ at every $\sig$ by
\be
N^0_\t(\sig)\mathbf{1}+iN^i_\t(\sig)\sig_i=g_{\sig\t}^{-}(g_{\sig\t}^{+})^{-1},\label{Ng}
\ee
where $\sig_i$ are Pauli matrices. 

\begin{Definition}\label{def0000}

A subclass $\mathscr{G}$ collects solutions $(g_{\sig\t}^\pm,\xi_{\t f})_c$ satisfying
\be
\sgn \lt[\det\lt(N_1(\sig),N_2(\sig),N_3(\sig),N_4(\sig)\rt)\rt]\neq 0.\label{nondeg4}
\ee
for all $\t=1,2,3,4$ out of 5. 

\end{Definition}

Note that due to gauge equivalence $g_{\sig\t}^\pm\sim \kappa_{\sig\t}^\pm h^\pm_{\sig} g_{\sig\t}^\pm$ ($\kappa_{\sig\t}^\pm=\pm 1$, $(h^+_{\sig},h^-_{\sig})\in\mathrm{Spin}(4)$) of $Z_\delta(\ck)$, five $N_\t(\sig)$s at $\sig$ are defined up to individual $\pm$ and a global SO(4) rotation.

We focus on solutions in the subclass $\mathscr{G}$. We construct at every $\sig$ 20 bivectors
\be
X_{\t f}(\sig)=(\vec{X}_{\t f}^+,\vec{X}_{\t f}^-)=\g J_{f}(\hat{g}_{\sig\t}^+\hat{n}_{\t f},\hat{g}_{\sig\t}^-\hat{n}_{\t f}),\label{bivector}
\ee
where $\hat{g}_{\sig\t}^\pm\in\mathrm{SO(3)}$ are 3-dimensional representations of ${g}_{\sig\t}^\pm$. $\vec{X}_{\t f}^\pm$ are self-dual and anti-self-dual parts: $X^\pm{}^i=\pm X^{0i}+\half\eps^i_{\ jk}X^{jk}$. 

Any 3 out of 4 $\hat{n}_{\t f}$ at every $\t$ is assumed to span a 3d space. In other words, it assumes that the tetrahedra reconstructed from the second equation in Eq.\Ref{critical} are all nondegenerate and
\be
\tr\lt(X_{\t f_1}[X_{\t f_2},X_{\t f_3}]\rt)(\sig)\neq 0,\quad \forall\ \t.\label{nondeg3}
\ee
Critical equations \Ref{critical} implies the following properties of $X_{\t f}(\sig)$: (1) $X_{\t f}(\sig)\wedge X_{\t f}(\sig)=0$, (2) $N_\t(\sig)\cdot X_{\t f}(\sig)=0$ for all $f\subset\t$, (3) $X_{\t f}(\sig)=X_{\t'f}(\sig)\equiv X_f(\sig)$ for all pairs of $\t,\t'\subset\sig$ with $\t\cap\t'=f$, and (4) $\sum_{f\subset\t} \eps_{\t f}(\sig)X_{\t f}(\sig)=0$. $\eps_{\t f}=\pm1$ satisfying $\eps_{\t f}(\sig)=-\eps_{\t' f}(\sig)$ ($\t\cap\t'=f$) and $\eps_{\t f}(\sig)=-\eps_{\t f}(\sig')$ ($\sig\cap\sig'=\t$). $\eps_{\t f}(\sig)$ is defined up a global sign on the entire $\ck$.

By Eqs.\Ref{nondeg4} and \Ref{nondeg3} and properties (1-4) of $X_{\t f}(\sig)$, the solution $(g_{\sig\t}^\pm,\xi_{\t f})_c$ (modulo gauge) reconstructs a unique 4-simplex geometry whose triangle areas are $\g J_f\ell_P^2$ on every $\sig\subset\ck$ \cite{semiclassicalEu}. Here each 4-simplex geometry is labelled by 10 edge lengths. Every geometrical 4-simplex gives 4d outward pointing normals $U_\t(\sig)$ to 5 boundary tetrahedra, such that $U_\t(\sig)$ satisfy a 4d closure condition and relate to the oriented 4-simplex volume
\be
\sum_{\t\subset\sig}U_\t(\sig)=0,\quad \frac{1}{V_\sig}=\det\lt(U_1,U_2,U_3,U_4\rt)(\sig)\label{UUUU}
\ee
The nondegeneracy $V_\sig\neq 0$ by Definition \ref{def0000}. $V_\sig$ relies on a consistent choice of ordering 4-simplex vertices (there is a 1-to-1 correspondence between vertices and tetrahedra in a 4-simplex), e.g. if $\sig=[1,2,3,4,5]$ with $\t=[1,2,3,4]$, a neighboring 4-simplex sharing $\t$ has to be $\sig'=-[1,2,3,4,5']$ inducing an opposite ordering to $\t$.

Geometrical 4d unit normals $\hat{U}_\t(\sig)=U_\t(\sig)/|U_\t(\sig)|$ are determined by the geometry up to global O(4) rotations at $\sig$. Relating to $N_\t(\sig)$ by $\hat{U}_\t(\sig)=\pm N_\t(\sig)$ reduces the ambiguity to global SO(4) rotations. There is also a gauge transformation on $g_{\sig\t}^\pm$ to set $\hat{U}_\t(\sig)=N_\t(\sig)$.

On the other hand, every geometrical 4-simplex give 20 bivectors $B_{\t f}(\sig)$ by
\be
B_{\t f}(\sig)&=&\g J_f*\frac{ \hat{U}_\t(\sig)\wedge \hat{U}_{\t'}(\sig)}{| \hat{U}_\t(\sig)\wedge  \hat{U}_{\t'}(\sig)|}\label{BUU}\\
&=& \half |V_\sig|*{U}_\t(\sig)\wedge {U}_{\t'}(\sig).\nonumber
\ee
The norm of a bivector $X$ is $|X|=\sqrt{\half X^{IJ}X_{IJ}}$. $B_{\t f}(\sig)$ relates to ``spinfoam bivectors'' $X_{\t f}(\sig)$ by\footnote{Here $\eps_{\t f}(\sig)X_{f}(\sig)$, $B_{\t f}(\sig)$ and $\mu(\sig)$ corresponds to $B_{ab}$, $B_{ab}(\sig)$ and $\mu$ in Barrett et al \cite{semiclassicalEu}.}
\be
\eps_{\t f}(\sig)X_{f}(\sig)&=&\mu(\sig) B_{\t f}(\sig)\nonumber\\
&=&\half\eps(\sig)\, V_\sig*{U}_\t(\sig)\wedge {U}_{\t'}(\sig),\label{XB}
\ee
where
\be
\eps(\sig)=\mu(\sig)\sgn(V_\sig)=\pm1.\label{muV}
\ee
$\mu(\sig)=1$ or $-1$ relates to that $\eps_{\t f}(\sig)\hat{n}_{\t f}$ are outward or inward pointing 3d face normals in all $\t\subset\sig$.\\

\noindent
\emph{2. Gluing 4-simplices:} Given neighboring $\sig,\sig'$ sharing $\t$, Eq.\Ref{bivector} implies $X_f(\sig)=(g^+_{\sig\sig'},g^-_{\sig\sig'})\cdot X_f(\sig')$ with $g^\pm_{\sig\sig'}=g^\pm_{\sig\t}g^\pm_{\sig'\t}{}^{-1}$. Then Eq.\Ref{XB} and $\eps_{\t f}(\sig)=-\eps_{\t f}(\sig')$ implies 
\be
B_{\t f}(\sig)=-\mu(\sig)\mu(\sig')(g^+_{\sig\sig'},g^-_{\sig\sig'})\cdot B_{\t f}(\sig').\label{BB}
\ee
$B_{\t f}(\sig)=\lt(\hat{g}_{\sig\t}^+\vec{b}_{\t f}(\sig),\hat{g}_{\sig\t}^-\vec{b}_{\t f}(\sig)\rt)$ where $\vec{b}_{\t f}(\sig)$ is the geometrical face normals of $\t$ from the 4-simplex geometry on $\sig$. $\vec{b}_{\t f}(\sig)$ satisfies the closure $\sum_{f\subset\t}\vec{b}_{\t f}(\sig)=0$ Eq.\Ref{BB} implies 
\be
\vec{b}_{\t f}(\sig)=-\mu(\sig)\mu(\sig')\vec{b}_{\t f}(\sig'),
\ee 
where the sign difference is independent of $f$. So tetrahedron geometries (labelled by edge lengths) from $\vec{b}_{\t f}(\sig)$ and $\vec{b}_{\t f}(\sig')$ coincide. Therefore 4-simplex geometries on $\sig,\sig'$ are glued with their induced tetrahedron geometries on $\t$ matching in shape. By gluing many 4-simplices to build $\ck$, the above shows that the solution $(g_{\sig\t}^\pm,\xi_{\t f})_c$ reconstructs a unique simplicial geometry labelled by edge lengths.  

\begin{Lemma}

$\eps(\sig)=\eps(\sig')=\eps$ for all $\sig,\sig'\subset\ck$, i.e. $\eps$ is a global sign on the entire $\ck$.

\end{Lemma}

\textbf{Proof:} Eq.\Ref{Ng} implies that $N_{\t}(\sig)=(g^+_{\sig\sig'},g^-_{\sig\sig'})\cdot N_{\t}(\sig')$ for $\t$ shared by $\sig,\sig'$, and implies 
\be
\hat{U}_{\t}(\sig)=s_{\sig\sig'}(g^+_{\sig\sig'},g^-_{\sig\sig'})\cdot \hat{U}_{\t}(\sig'),\quad s_{\sig\sig'}=\pm1
\ee
where $s_{\sig\sig'}$ comes from the sign gauge ambiguity relating $\hat{U}_\t$ and $N_\t$. Moreover by $X_f(\sig)=(g^+_{\sig\sig'},g^-_{\sig\sig'})\cdot X_f(\sig')$,
\be
\eps_{\t f}(\sig)X_{f}(\sig)&=&\half\eps(\sig)\, V_\sig*{U}_\t(\sig)\wedge {U}_{\t_1}(\sig)\nonumber\\
&=&-\half\eps(\sig')\, V_{\sig'}*{U}'_\t(\sig)\wedge {U}'_{\t_1'}(\sig),
\ee 
where ${U}'_\t(\sig)=(g^+_{\sig\sig'},g^-_{\sig\sig'})\cdot U_\t(\sig')$. Since ${U}'_\t(\sig)\propto {U}_\t(\sig)$, ${U}'_{\t_1'}(\sig)$ is a linear combination of ${U}_\t(\sig),{U}_{\t_1}(\sig)$. Explicitly
\be
{U}'_{\t_1'}(\sig)=-s_{\sig\sig}\frac{\eps(\sig)\,|U_\t(\sig)|\,V_\sig}{\eps(\sig')\,|U_\t(\sig')|\,V_{\sig'}}{U}_{\t_1}(\sig)+ a_1 {U}_{\t}(\sig).\label{UUU}
\ee 
$V_\sig,V_{\sig'}$ are given by
\be
V_\sig^{-1}&=&\det\lt({U}_{\t_1},{U}_{\t_2},{U}_{\t_3},{U}_{\t_4}\rt)(\sig)\nonumber\\
V_{\sig'}^{-1}&=&-\det\lt({U}'_{\t_1'},{U}'_{\t_2'},{U}'_{\t_3'},{U}'_{\t_4'}\rt)(\sig)\label{VUUUU}
\ee
since $\det$ of $U$s is invariant under SO(4) rotations. The minus sign comes from the ordering $\sig=[1,2,3,4,5]$ and $\sig'=-[1,2,3,4,5']$. Eq.\Ref{UUU} is also valid to ${U}'_{\t_2'},{U}'_{\t_3'},{U}'_{\t_4'}$. Because $\sum_{\t'}U'_{\t'}=0$,
\be
V_{\sig'}^{-1}&=&-\det\lt({U}'_\t, {U}'_{\t_1'},{U}'_{\t_2'},{U}'_{\t_3'}\rt)(\sig)\label{123}\\
&=& \frac{\eps(\sig)}{\eps(\sig')}\lt(\frac{|U_\t(\sig)|\,V_\sig}{|U_\t(\sig')|\,V_{\sig'}}\rt)^2V_{\sig'}^{-1}
\ee
which implies $\eps(\sig)=\eps(\sig')$ and $|U_\t(\sig)|\,V_\sig=\pm |U_\t(\sig')|\,V_{\sig'}$. $\Box$ \\

The appearance of global sign ambiguity $\eps$ comes from critical equation Eq.\Ref{critical} is invariant under a global refection $\hat{n}_{\t f}\to -\hat{n}_{\t f}$ or $\xi_{\t f}\to J\xi_{\t f}$ on the entire $\ck$ (named ``global $J$-parity'' in \cite{hanPI}). But this invariance is broken when $\ck$ has a boundary where some $\hat{n}_{\t f}$s are fixed by the boundary condition. In this case, we can set e.g. $\eps=1$ by redefining $\eps_{\t f}(\sig)$ globally. If $\ck$ has no boundary, $\eps=\pm1$ corresponds to 2 different solutions related by this global refection of $\hat{n}_{\t f}$.

When $\eps(\sig)=\eps=1$, Eq.\Ref{muV} gives
\be
\mu(\sig)=\sgn(V_\sig).
\ee

The above proves the forward direction in the correspondence Eq.\Ref{equiv0}:

\begin{Theorem}

Given any solution $(g_{\sig\t}^\pm,\xi_{\t f})_c\in\mathscr{G}$ (modulo gauge) to critical equations \Ref{critical}, it reconstructs uniquely a nondegenerate simplicial geometry labelled by edge lengths on $\ck$, and determines all 4-simplex orientations $\sgn(V_\sig)=\pm1$, which is not constant in general. The solution also give a global sign $\eps=1$ or $-1$ when $\partial\ck=\emptyset$.

\end{Theorem}

The reconstruction defines a map 
\be
\cc:\ \mathscr{G} \to \text{the space of $(\ell,\sgn(V_\sig),\eps)$},
\ee 
where $\ell$ labels a simplicial geometry on $\ck$, and $\sgn(V_\sig)$ labels the 4-simplex orientation. The following discusses the injectivity and surjectivity of $\cc$.\\

\noindent
\emph{3. Injectivity and surjectivity of $\cc$:} Given data $(\ell,\sgn(V_\sig),\eps)$ where $\ell$ is a nondegenerate simplicial geometry on $\ck$ with edge lengths $\ell$ and triangle areas $\g J_f(\ell)$, $\sgn(V_\sig)$ are orientations at all $\sig$s, and $\eps$ is a global sign ($\eps=-1$ when $\ck$ has boundary), we suppose that $(\ell,\sgn(V_\sig),\eps)$ can be reconstructed by 2 different solutions $(g_{\sig\t}^\pm,\xi_{\t f})_c,\, (g'{}_{\sig\t}^\pm,\xi'_{\t f})_c \in\mathscr{G}$.

$(\ell,\sgn(V_\sig),\eps)$ determines 4d unit normals ${U}_\t(\sig)$ outward pointing from every $\sig$, up to global SO(4) rotations at $\sig$ by Eq.\Ref{UUUU}. We set $\hat{U}_\t(\sig)=N_{\t}(\sig)=\hat{G}_{\sig\t} \cn$ where $\hat{G}_{\sig\t}\in \mathrm{SO(4)}$ and $\cn=(1,0,0,0)$. Individual $\hat{G}_{\sig\t}$s are fixed by this relation up SO(3) rotations leaving $\cn$ invariant. Up to this SO(3), $\hat{G}_{\sig\t}$ is the 4-dimensional representation of both $g_{\sig\t}^\pm$ and $g'{}_{\sig\t}^\pm$ up to gauge freedom. $\hat{U}_\t(\sig)=N_{\t}(\sig)$ fixes the discrete gauge freedom of $g_{\sig\t}^\pm$ up to $g_{\sig\t}^\pm\to \kappa_{\sig\t}g_{\sig\t}^\pm$, $\kappa_{\sig\t}=\pm1$ leaving $N_\t(\sig)$ invariant.

The geometrical bivectors $B_{\t f}(\sig)$ given by ${U}_\t(\sig)$ in Eq.\Ref{BUU} and acted by $\hat{G}_{\sig\t}^{-1}$ gives a bivector orthogonal to $\cn$:
\be
\hat{G}_{\sig\t}^{-1} B_{\t f}(\sig)=\lt(\vec{b}_{\t f}(\sig),\vec{b}_{\t f}(\sig)\rt),\quad \lt|\vec{b}_{\t f}^\pm(\sig)\rt|=\g J_f(\ell)
\ee  
A set of 4 3d vectors $\vec{b}_{\t f}(\sig)$ and $\vec{b}_{\t f}(\sig')$ are related by an SO(3) rotation leaving $\cn$ invariant, because both of them are face normals of a geometrical tetrahedron shared by $\sig,\sig'$. So we can implement this SO(3) rotation to $\hat{G}_{\sig\t}$ or $\hat{G}_{\sig'\t}$ to make
\be
\vec{b}_{\t f}(\sig)=\pm \vec{b}_{\t f}(\sig').\label{bb00}
\ee
This reduces ambiguities of $\hat{G}_{\sig\t}$ and $\hat{G}_{\sig'\t}$ from SO(3)$\times$SO(3) to SO(3): $\hat{G}_{\sig\t},\, \vec{b}_{\t f}(\sig) \sim \hat{G}_{\sig\t}\hat{h}_\t,\, \hat{h}_\t^{-1}\vec{b}_{\t f}(\sig) $ where $\hat{h}_\t\in \mathrm{SO(3)}$ independent of $\sig$.

\begin{Lemma}\label{VbVb}

$\vec{b}_{\t f}(\sig)=-\sgn(V_\sig)\,\sgn(V_{\sig'}) \,\vec{b}_{\t f}(\sig')$ is implied by $(\ell,\sgn(V_\sig),\eps)$. 

\end{Lemma}

\textbf{Proof:} $\hat{G}_{\sig\t}^{-1} B_{\t f_i}(\sig)=*\cn\wedge\vec{b}_{\t f_i}(\sig)$ so $\hat{G}_{\sig\t}^{-1} \hat{U}_{i}\propto\vec{b}_{\t f_i}(\sig)+ \a_i\cn$ for $\t_{i=1,2,3,4}\subset\sig$ sharing $f_i$ with $\t$. Eq.\Ref{UUUU} implies that $\hat{U}_\t,\ \sgn(V_\sig)\hat{U}_1,\  \sgn(V_\sig)\hat{U}_2,\ \sgn(V_\sig)\hat{U}_3$ form a right-hand frame at $\sig$. Rotating by $\hat{G}_{\sig\t}\in\mathrm{SO(4)}$ implies that 
\be
\cn,\ \sgn(V_\sig)\vec{b}_{\t f_1}(\sig),\  \sgn(V_\sig)\vec{b}_{\t f_2}(\sig),\ \sgn(V_\sig)\vec{b}_{\t f_3}(\sig)\label{righthandf}
\ee 
form a right-hand frame. By Eqs.\Ref{VUUUU} and \Ref{123}, from $\sig'$ we obtain the right-hand frame 
\be
\cn,\ -\sgn(V_{\sig'})\vec{b}_{\t f_1}(\sig'),\  -\sgn(V_{\sig'})\vec{b}_{\t f_2}(\sig'),\ -\sgn(V_{\sig'})\vec{b}_{\t f_3}(\sig').\nonumber
\ee 
By Eq.(\ref{bb00}) and comparing to \Ref{righthandf}, we obtain
\be
\sgn(V_\sig)\vec{b}_{\t f}(\sig)=-\sgn(V_{\sig'})\vec{b}_{\t f}(\sig').
\ee
$\Box$\\

Lemma \ref{VbVb} is consistent with Eq.\Ref{XB} which implies 
\be
\eps\,\sgn(V_\sig)\,\eps_{\t f}(\sig) \,\vec{b}_{\t f}(\sig)=\g J_f(\ell)\, \hat{n}_{\t f}
\ee
since $\hat{G}_{\sig\t}$ is the 4-dimensional representation of $g_{\sig\t}^\pm$ or $g'{}_{\sig\t}^\pm$. It determines $\hat{n}_{\t f}$ up to $\hat{h}_\t\in \mathrm{SO(3)}$.

As a result, $(\ell,\sgn(V_\sig),\eps)$ determines $(\hat{G}_{\sig\t},\hat{n}_{\t f})$ up to gauge freedom $(\hat{G}_{\sig\t},\, \hat{n}_{\t f}) \sim (\hat{h}_\sig\hat{G}_{\sig\t}\hat{h}_\t,\, \hat{h}_\t^{-1}\hat{n}_{\t f})$ with $\hat{h}_\sig\in \mathrm{SO(4)}$ and $\hat{h}_\t\in \mathrm{SO(3)}$. Therefore modulo the gauge freedom, $g_{\sig\t}^\pm$ and $g'{}_{\sig\t}^\pm$ are 2 different lifts from $\hat{G}_{\sig\t}\in \mathrm{SO(4)}$ to Spin(4), thus $g_{\sig\t}^\pm= \kappa_{\sig\t}g'{}_{\sig\t}^\pm$, $\kappa_{\sig\t}=\pm1$. But $g_{\sig\t}^\pm\to \kappa_{\sig\t}g_{\sig\t}^\pm$ is a discrete gauge transformation of the SFM. Moreover $\hat{n}_{\t f}$ determines that $\xi_{\t f}=e^{i\theta_{\t f}}\xi'_{\t f}$, while $\xi_{\t f}\to e^{i\theta_{\t f}}\xi_{\t f}$ is again a gauge transformation for internal $\xi_{\t f}$, and the phase ambiguity of $\xi_{\t f}$ at boundary is fixed in any boundary condition. Therefore $(g_{\sig\t}^\pm,\xi_{\t f})_c=(g'{}_{\sig\t}^\pm,\xi'_{\t f})_c$ modulo continuous and discrete gauge transformations. Nondegeneracy of the simplicial geometry $\ell$ implies that $(g_{\sig\t}^\pm,\xi_{\t f})_c\in\mathscr{G}$.

The above proves that the map $\cc$ is injective. It also proves $\cc$ is surjective because we start from arbitrary data $(\ell,\sgn(V_\sig),\eps)$ and recover a solution $(g_{\sig\t}^\pm,\xi_{\t f})_c\in\mathscr{G}$.

\begin{Theorem}

The map $\cc$ relating solutions $(g_{\sig\t}^\pm,\xi_{\t f})_c\in\mathscr{G}$ to nondegenerate simplicial geometries and orientations $(\ell,\sgn(V_\sig),\eps)$ is a bijection. 

\end{Theorem}

\subsection{Deficit Angles}\label{angles}

Given a critical solution $X_c\equiv (g_{\sig\t}^\pm,\xi_{\t f})_c\in\mathscr{G}$ corresponding to $ (\ell,\sgn(V_\sig),\eps)$ with $\sgn(V_{\sig})=1$ at all $\sig$ and $\eps=1$, $F_f$ evaluated at $X_c$ gives \cite{CFsemiclassical,HZ1,Han:2017xwo},
\be
F_f[X_c] =i\lt(\Phi_f^++\Phi_f^-\rt)+i\g\lt(\Phi_f^+-\Phi_f^-\rt),\quad \g=p/q
\ee
where $p/q\in\Z_+$ and $p+q$ is an even number.
\be
\Phi_f^\pm&=&\sum_{\sig,f\subset\sig}\phi^\pm_{\t\sig\t'},\nonumber\\
i\phi^\pm_{\t\sig\t'}&=&\ln\lag{\xi_{\t f}}\big|(g^\pm_{\sig\t})^{-1}g^\pm_{\sig\t^\prime}\big|{\xi_{\t^\prime f}}\rag\Big|_{X_c}\in i\R
\ee
Recall that the integrand in $Z_\delta(\ck)$ depends on $F_f$ through $F_f+4\pi i k_f/q +\pi i\sum_{\pm}m_\t\frac{q\pm p}{q}$, and $Z_\delta(\ck)$ sums $k_f,m_\t\in\Z$. The integrand in $Z_\delta(\ck)$ is invariant under the following shifts:
\be
\Phi_f^++\Phi_f^-\to \Phi_f^++\Phi_f^-+4\pi,  &\text{or}& \Phi_f^+-\Phi_f^-\to \Phi_f^+-\Phi_f^-+4\pi,\nonumber\\
\text{and}\quad k_f\to k_f- q,  &\text{or}& k_f\to k_f- p
\ee
and 
\be
\Phi_f^++\Phi_f^-\to \Phi_f^++\Phi_f^-+2\pi,  &\text{and}&  \Phi_f^+-\Phi_f^-\to \Phi_f^+-\Phi_f^-+2\pi,\nonumber\\
\text{and}\quad k_f\to k_f-(q+p)/2.&&
\ee
The above gauge invariance allows us to fix the following range of angles:
\be
\Phi_f^++\Phi_f^-\in[-2\pi,2\pi],\quad \Phi_f^+-\Phi_f^-\in[-\pi,\pi].\label{B3}
\ee

\begin{figure}[htbp]
	\centering\includegraphics[width=0.4\linewidth]{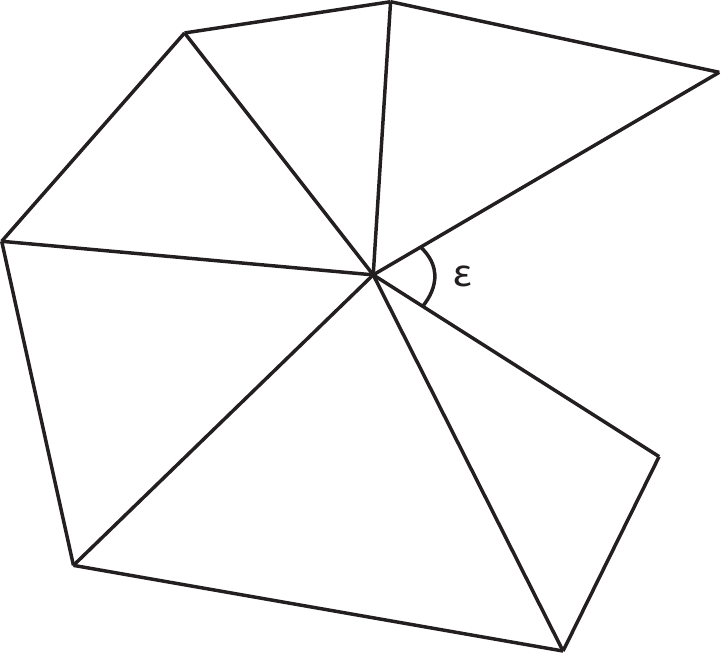}
	\caption{The deficit angle $\eps$ in a 2d discrete surface hinged by a point. $\eps\neq0$ demonstrates that summing the angles at the hinge fails to give $2\pi$. One obtains a discrete curved surface when the two edges bounding $\eps$ are glued. In higher dimensions, $\eps$ is always hinged by a co-dimension-2 simplex, e.g. in 4d, $\eps_f$ is hinged by a triangle $f$. }\label{fig:DeficitAngle}
\end{figure}

At $X_c$, $\phi^\pm_{\t\sig\t'}$ relates to the 4d dihedral angle $\theta_{f}(\sig)$ between the two tetrahedra $\t$ and  $\t'$ within $\sig$.: \cite{semiclassicalEu}
\be
\phi^+_{\t\sig\t'}-\phi^-_{\t\sig\t'}=\pi-\theta_{f}(\sig)\in[0,\pi].\label{phiphi}
\ee
We define $n_f$ to be the number of $\sig$ sharing an internal $f$. $n_f$ is always even for the triangulation $\ck$ adapted to a hypercubic lattice (see Appendix \ref{TPOT}). Then shifting by multiples of $2\pi$ and $4\pi$ gives
\be
\Phi_f^+-\Phi_f^-&=&n_f\pi-\sum_{\sig, f\subset\sig}\theta_f(\sig)-4\pi u-2\pi v\nonumber\\
&=&2\pi-\sum_{\sig, f\subset\sig}\theta_f(\sig)=\eps_f
\ee
for  certain $u,v\in\Z$. The deficit angle $\eps_f$ hinged by $f$ is a discrete description of Riemann curvature in simplicial geometry (FIG.\Ref{fig:DeficitAngle}). 

To determine $\Phi_f^++\Phi_f^-$, we consider all $g_{\sig\t}^\pm$ whose $\sig$s and $\t$s share a single internal triangle $f$. At the solution, $g^\pm_{\t\sig} g^\pm_{\sig\t'}|\xi_{\t' f}\rangle= e^{i\phi^\pm_{\t\sig\t'}}|\xi_{\t f}\rangle$ where $g^\pm_{\t\sig}= (g^\pm_{\sig\t})^{-1}$, so
\be
g_{\t \sig_1}^{\pm }g_{\sig_{1}\t_{1}}^{\pm }\cdots g_{\t_{k}\sig}^{\pm }g_{\sig \t}^{\pm}|\xi_{\t f}\rangle= e^{i\Phi_f^\pm}|\xi_{\t f}\rangle,
\ee
which gives 
\be
g_{\t \sig_1}^{\pm }g_{\sig_{1}\t_{1}}^{\pm }\cdots g_{\t_{k}\sig}^{\pm }g_{\sig \t}^{\pm}=g(\xi_{\t,f})\begin{pmatrix}
  e^{i\Phi_f^\pm} & 0\\
  0& e^{-i\Phi_f^\pm}
\end{pmatrix} g(\xi_{\t,f})^{-1}
\ee
where $g(\xi)=(\xi,J\xi)\in\Su$. We define
\be
G_f^\pm(\sig)\equiv g_{\sig \t}^{\pm}g_{\t \sig_1}^{\pm }g_{\sig_{1}\t_{1}}^{\pm }\cdots g_{\t_{k}\sig}^{\pm }=\exp \left( i\Phi _{f}^{\pm }\hat{\bf X}_{f}^{\pm}(\sig)\right) \label{Gf}
\ee
where $\Phi _{f}^{\pm }\sim \Phi _{f}^{\pm }+ 2\pi$ and 
\be
\hat{\bf X}_{f}^{\pm}(\sig)=g^\pm_{\sig\t}g(\xi_{\t,f})\sig_3g(\xi_{\t,f})^{-1}(g_{\sig\t}^\pm)^{-1}=g^\pm_{\sig\t}(\hat{n}_{\t f}\cdot\vec{\sig})(g_{\sig\t}^\pm)^{-1}. \nonumber
\ee
Comparing to Eq.\Ref{bivector}, $\hat{\bf X}_{f}^{\pm}(\sig)=\vec{X}_{f}^{\pm}(\sig)\cdot\vec{\sig}/|\vec{X}_{f}^{\pm}(\sig)|$.

On the other hand, in terms of the corresponding geometry,
\be
E_{jk}(\sig)={V_\sig} * U_{l}(\sig)\wedge U_{m}(\sig)\wedge U_{n}(\sig)
\ee
defines an edge vector $(|E_{jk}(\sig)|=\ell_{jk})$ pointing to the vertex $j$ from the vertex $i$ \cite{CFsemiclassical,HZ1}. Here $\sig=[k,j,l,m,n]$ is the ordering of vertices. Eq.\Ref{UUU} implies that for all edges of $\t$ $j,k=1,\cdots,4$
\be
(g^+_{\sig\sig'},g^-_{\sig\sig'})\cdot E_{jk}(\sig')=s_{\sig\sig'}\sgn(V_\sig)\sgn(V_{\sig'})E_{jk}(\sig).
\ee
We have assumed $\sgn(V_\sig)=1$ at all $\sig$, and we partially gauge fix the discrete gauge freedom of $g^\pm_{\sig\t}$ such that $\hat{U}_\t(\sig)=N_\t(\sig)$ so $s_{\sig\sig'}=1$ (the remaining discrete gauge freedom is $g_{\sig\t}^\pm\to \kappa_{\sig\t}g_{\sig\t}^\pm$, $\kappa_{\sig\t}=\pm1$). So $(g^+_{\sig\sig'},g^-_{\sig\sig'})$ is a discrete spin connection. By the parallel transport of $E_{jk}(\sig)$, 
\be
\hat{G}_f(\sig)E_{jk}(\sig)=E_{jk}(\sig), \quad \forall (j,k)\subset f.
\ee
Therefore $\hat{G}_f(\sig)\in\mathrm{SO(4)}$ (the 4-dimensinal representation of $G_f(\sig)$) is a 4d rotation leaving the geometrical triangle $f$ invariant
\be
\hat{G}_f(\sig)=\exp\left( -*\hat{X}_{f}(\sig)\,\vartheta_f\right)\label{hatGf}
\ee
where by $V_\sig * U_m(\sig)\wedge U_{n}(\sig)=E_{jk}(\sig)\wedge E_{lk}(\sig)$,
\be
\eps_{\t f}(\sig)\hat{X}_{f}(\sig)=\frac{\eps_{\t f}(\sig) X_f(\sig)}{|X_f(\sig)|}=\frac{B_{\t f}(\sig)}{|B_{\t f}(\sig)|}=\frac{E_{jk}(\sig)\wedge E_{lk}(\sig)}{|E_{jk}(\sig)\wedge E_{lk}(\sig)|}.\nonumber
\ee

$(G_f(\sig)^+,G_f^-(\sig))\in\mathrm{Spin(4)}$ in Eq.\Ref{Gf} is a lift of $\hat{G}_f(\sig)\in\mathrm{SO(4)}$ in Eq.\Ref{hatGf}. The angles $\Phi_f^\pm$ and $\vartheta_f$ are related by
\be
\Phi^{+}_f-\Phi_f^{-}=\vartheta_f,\quad \Phi^{+}_f+\Phi_f^{-}=2\chi_f\pi,\quad \chi_f\in\{0,1\}
\ee
or $\Phi^{\pm}_f=\pm\frac{1}{2}\vartheta_f-\chi_f\pi$, where $\chi_f$ labels the lift ambiguities from SO(4) to Spin(4). Note that similar as above, the periodicity $\Phi^{\pm}_f\sim \Phi^{\pm}_f+2\pi$ allows us to set $\vartheta_f\in[-\pi,\pi]$ and $\chi_f\in\{0,1\}$. This identifies $\vartheta_f=\eps_f$.

In general most lift ambiguities can be canceled by the remaining discrete gauge freedom $g_{\sig\t}^\pm\to \kappa_{\sig\t}g_{\sig\t}^\pm$, $\kappa_{\sig\t}=\pm1$:

\begin{Lemma}\label{coho}

The lift ambiguities $\chi_f$ at all internal $f$ are removed by discrete gauge transformations $g_{\sig\t}^\pm\to \kappa_{\sig\t}g_{\sig\t}^\pm$, $\kappa_{\sig\t}=\pm1$ up to $H^2(\ck^*,\Z_2)$, the 2nd cellular cohomology group of the dual complex $\ck^*$.

\end{Lemma}

\textbf{Proof:} A spin structure on the manifold triangulated by $\ck$ defines a canonical lift of $\hat{G}_{f}(\sig)$ to $(\O^+_f(\sig),\O^-(\sig))\in\mathrm{ Spin(4)}$ such that the lift $\O^\pm_f(v)$ can be continuously deformed to $1$. ${G}_f^\pm(\sig)=e^{i\pi \chi_f}\O^\pm_f(v)$ where $\chi_f\in\{0,1\}\simeq \Z_2$ gives the other lift of $\hat{G}_f^\pm(\sig)$ when $\chi_f=1$. 

Given a triangulated manifold $\ck$, there is a corresponding dual polyhedral decomposition $\ck^*$. Given an edge $\ell$ shared by a number of internal $f$ in $\ck$. $\ell$ is dual is a 3d polyhedron $\ell^*$ bounded by internal faces $f^*$s dual to $f$s in $\ck^*$. ${G}_f^\pm(\sig)$ and $\O^\pm_f(\sig)$ are along the dual face boundary $\partial f^*$, and based at the dual vertex $\sig^*$. The polyhedron $\ell^*$ gives cocycle conditions to both ${G}_f^\pm, {\O}_f^\pm$: 
\be
\overrightarrow{\prod_f}{\O}_f^\pm=1\quad\text{and} \quad\overrightarrow{\prod_f}{G}_f^\pm=1,\label{Gcocycle}
\ee 
where all ${\O}_f^\pm$s (${G}_f^\pm$s) are parallel transported by $\O^{\pm}_{\sig\sig'}$ ($G^{\pm}_{\sig\sig'}$) to share the same base point. The above relations may be seen by viewing ${\O}_f^\pm$s (${G}_f^\pm$s) are flat connection holonomies on a 2-sphere with $p$ holes (each ${\O}_f^\pm$ (${G}_f^\pm$) circles around a hole), followed by enlarging holes to approach the skeleton of the polyhedron $\ell^*$ with $p$ faces.

Parallel transports are made by conjugate with $\O^{\pm}_{\sig\sig'}$ or $G^{\pm}_{\sig\sig'}$ whose sign ambiguities doesn't affect $e^{i\pi \chi_f}$. Eqs.\Ref{Gcocycle} result in the $\Z_2$-cocycle condition
\be
\sum_{f} \chi_f=0,\quad \chi_f\in\Z_2
\ee 
If we understand $\chi_f=\lag f^*,\chi_2\rag$ where $\chi_2$ is a 2-cochain, then $\sum_{f} \chi_f=\langle \partial \ell^*,  \chi_2\rangle=\langle \ell^*, {\delta} \chi_2\rangle $ i.e. $\delta \chi_2=0$ where $\delta$ is the coboundary differential. 

If $\ck^*$ has a nontrivial 2nd cohomology goup $H^2(\ck^*,\Z_2)$, there exist $\eta\in H^2(\ck^*,\Z_2)$, such that $\chi_2=\eta+\delta \chi_1$. 
Evaluate at any dual face $f^*$ gives
\be
\chi_f= \lag f^*,\eta\rag+\lag \partial f^*,  \chi_1\rag= \lag f^*,\eta\rag+\sum_\t\chi_{\t}, \quad \chi_{\t}\in\Z_2,\label{chichi}
\ee
where $\chi_{\t}=\langle \t^*,\chi_1\rangle$ and $\sum_\t$ is over all $\t^*\subset\partial f^*$. Eq.\Ref{chichi} implies there exists $\chi_\t\in\Z_2$ such that
\be
e^{i\pi\chi_f}=e^{i\pi \lag f^*,\eta\rag}\prod_\t e^{i\pi\chi_\t},
\ee
The factor $\prod_\t e^{i\pi\chi_\t}$ can be cancelled by the discrete gauge transformation $g_{\sig\t}^\pm\to e^{i\pi\chi_\t} g_{\sig\t}^\pm$ at one $\sig$ bounded by $\t$. Therefore we obtain
\be
G^\pm_{f}(\sig)=e^{i\pi \lag f^*,\eta\rag}\O^\pm_{f}(\sig).\label{}
\ee
where $\eta\in H^2(\ck^*,\Z_2)$. $\Box$\\

When $\ck^*$ is a polyhedral decomposition of $\R^4$ as in our main discussion, all lift ambiguity can be removed by gauge transformations since $H^2(\R^4,\Z_2)=0$. When $\ck^*$ has boundary and is a polyhedral decomposition of a (topologically trivial) compact 4d region $\calr\subset\R^4$, we apply Lefschetz-Poincar\'e duality $H^2(\calr,\Z_2)\simeq H_2(\calr,\partial\calr,\Z_2)=0$ where $H_2(\calr,\partial\calr,\Z_2)$ is the 2nd relative homology. Since Lemma \ref{coho} is valid only for internal $f$s, gauge transformations may not be able to remove lift ambiguities at boundary $f$s. $i(\Phi_f^++\Phi^-)=i\pi$ may present in the boundary $F_f[X_c]$. Boundary $F_f$'s doesn't affect our derivation of Eqs.\Ref{eom} and \Ref{small}.

As a result, we conclude that when $H^2(\ck^*,\Z_2)=0$, for all internal $f$, 
\be
F_f[X_c]=i\g \eps_f. 
\ee

\section{Background and perturbations}\label{BAP}

We define the background $\mathring{X}$ as the solution in the subclass $\mathscr{G}$, corresponding (as Eq.\Ref{equiv0}) to the flat simplicial geometry whose edge-lengths are $\mathring{\ell}$ on $\ck$, and with a uniform 4-orientation at all 4-simplices $\sgn(V_\sig)=1$. Recall $\mathring{\ell}$ is a flat triangulated hypercubic lattice with constant spacing $(\g\l)^{1/2}\ell_P$. The geometry and orientation uniquely fixes the critical point $\mathring{X}$ by Eq.\Ref{equiv0}.

When we perturb $\mathring{X}\in\mathscr{G}$ by $(\mathring{J}+\delta J,\mathring{X}+\delta X)$ where $\mathring{J}=J(\mathring{\ell})$. There is a neighborhood at $\mathring{X}$ such that all other solutions $X_c\neq \mathring{X}$ still belong to the subclass $\mathscr{G}$, and have the same uniform orientation $\sgn(V_\sig)=1$ for all $\sig$. Note that here $X_c$ may associates to a different $\mathring{J}+\delta J$. Indeed $\mathring{X}$ at every $\sig$ gives $(\mathring{g}_{\sig\t}^+,\mathring{g}_{\sig\t}^-)$ and $\mathring{\xi}_{\t f}$ with $\mathring{g}_{\sig\t}^+\neq\mathring{g}_{\sig\t}^-$. Here $\mathring{g}_{\sig\t}^+, \mathring{g}_{\sig\t}^-$ are very different, or namely there is a finite distance between $\mathring{g}_{\sig\t}^+, \mathring{g}_{\sig\t}^-\in\Su$ measured by the natural metric on $S^3$, because $N_{\t}(\sig)$ at 5 $\t\subset\partial\sig$ determined by $(\mathring{g}_{\sig\t}^+,\mathring{g}_{\sig\t}^-)$ are far from being parallel. Therefore there exists a neighborhood at $(\mathring{g}_{\sig\t}^+,\mathring{g}_{\sig\t}^-)$, such that perturbations $(\mathring{g}_{\sig\t}^++\delta {g}_{\sig\t}^+,\mathring{g}_{\sig\t}^-+\delta {g}_{\sig\t}^-)$ only perturb $N_\t(\sig)$ but do not change $\sgn(V_\sig)$. Perturbations neither interchange $\mathring{g}_{\sig\t}^+,\mathring{g}_{\sig\t}^-$ nor make them equal. As a result, perturbations $(\mathring{J}+\delta J,\mathring{X}+\delta X)$ can only touch solutions in $\mathscr{G}$ having the same uniform orientation as $(\mathring{J},\mathring{X})$, but cannot touch solutions with different orientations. Perturbations cannot touch BF-type solutions ($g^+_{\sig\t},g^+_{\sig\t}$), ($g^-_{\sig\t},g^-_{\sig\t}$) and vector geometry solutions ($g_{\sig\t},g_{\sig\t}$) because their $N_\t(\sig)$s are all parallel and give $\sgn(V_\sig)=0$. Therefore the existences of cosine problem and BF-type, vector geometry solutions do not affect our following derivation of perturbative solutions, since no solution of opposite oriention, BF-type, or vector geometry appears in perturbations. We focus in the following on $X_c\in\mathscr{G}$ with the same orientation as $\mathring{X}$.

Note the above argument doesn't work for SU(2) BF theory: The critical equation provides only the $+$ sector of Eq.\Ref{critical}, so the subclass $\mathscr{G}$ doesn't exist in the BF theory. Solutions there only contain BF-type solutions of nondegenerate geometries, and vector geometries. But BF-type solutions are not well-separated from vector geometry solutions \cite{DFSS}. Therefore perturbations from a nondegenerate geometry can touch vector geometries. 

By the correspondence Eq.\Ref{equiv0}, and since the orientation is preserved by perturbations, all perturbative solutions from critical equations $\mathrm{Re} S=\delta_X S=0$ or Eq.\Ref{critical} in $(\mathring{J}+\delta J,\mathring{X}+\delta X)$ correspond to perturbations of simplicial geometries with edge lengths $\ell=\mathring{\ell}+\delta\ell$. i.e. the geometries are perturbations of the flat simplicial geometry $\mathring{\ell}$ on $\ck$.

Any solution $X_c\in\mathscr{G}$ with a uniform orientation as $\mathring{X}$ implies ${F}_f[X_c]=i\g\eps_f(\ell)$ (see Subsection \ref{angles} for a proof) where $\eps_f$'s are deficit angles, which measure discrete Riemannian curvature. This applies in particular to the above perturbative solutions.

\section{Equation of motion and small deficit angles}\label{EOMASDA}

In the above, we have obtained the perturbative solutions of a subset of critical equations $\mathrm{Re} S=\delta_X S=0$ and their geometrical interpretations. The other critical equation $\delta_\ell S=0$ and Eq.\Ref{area-length} yields the equation of motion (EOM)
\be
\lag \frac{\partial\vec{J}(\ell)}{\partial\ell},\g\vec{\eps}(\ell)\rag=0\ \ \text{or}\ \ \sum_f\frac{\partial\mathbf{a}_f(\ell)}{\partial\ell}{\eps}_f(\ell)=0,\label{eom}
\ee
and coincides with the Regge equation. Regge equation is a discretization of Einstein equation in 4d \cite{regge}.

The leading asymptotic behavior of Eq.\Ref{Z1} is determined by the integrand evaluated at the critical point:
\be
e^{S_{eff}}\Big|_{\text{critical\ pt}}= e^{i\lag \g\vec{J}(\ell),\vec{\eps}(\ell)\rag-\sum_{i=1}^M\frac{1}{\delta}\langle \hat{e}^i,\g\vec{\eps}(\ell)\rangle^2}.\label{Rcorrrection}
\ee
The first term in the exponent is the Regge action which vanishes at the solution of Eq.\Ref{eom}. The second term is the $\eps_f^2$ higher curvature correction (mentioned in the last section) which encodes the contributions from non-Regge-like $J$s. Since the $\eps_f^2$ correction term is real and negative, and $\delta\ll1$, it suppresses the contribution of the critical point $(\ell_c,X_c)$ exponentially unless $\lt|\lag \hat{e}^i,\g\vec{\eps} \rag\rt|\lesssim \delta^{1/2}$ for all $i$. Since  $\{{\partial\vec{J}(\ell)}/{\partial\ell},\hat{e}^i\}$ forms a complete basis in $\R^{N_f}$, it follows from  Eq.\Ref{eom} that
\be
|\g\eps_f(\ell)|\lesssim\delta^{1/2}\ll1,\label{small}
\ee
Eqs. \Ref{eom} and \Ref{small} determine the critical points $(\ell_c,X_c)$ that contribute essentially to $Z_\delta(\ck)$, and thus are the key equations constraining the simplicial geometries emerging in the large spin limit of the model.


Eqs.\Ref{eom} and \Ref{small} are trivially satisfied by the flat background $(\mathring{J},\mathring{X})$. $(\mathring{J},\mathring{X})$ is a critical point of $Z_\delta(\ck)$ thus is qualified as a background. For perturbations, Eq.\Ref{eom} can be reduced to a set of linear equations of the deficit angles $\eps_f$ \cite{0264-9381-5-12-007}, because the considered geometries are nearly-flat. That is, 
\be
{M}\vec{\eps}=0. \label{lin} 
\ee 
where $M$ is a constant $N_f\times N_f$ matrix. Note that this is a consequence of the nearly-flat geometries, but not a consequence of Eq.\Ref{small}. By itself, Eq.\Ref{small} is compatible with the non-linear Regge equation, and excludes no nonsingular curved geometry. 
On a sufficiently refined triangulation, any simplicial geometry approximating a smooth geometry with typical curvature radius $\rho$ satisfies $|\eps_f|\simeq a^2/\rho^2\ll1$, which is consistent with Eq.\Ref{small}. Here $a$ is the typical lattice spacing. 
The simplicial geometries that fail to satisfy Eq.\Ref{small} cannot have smooth approximation.

If the regularization in Eq.\Ref{mod} wasn't imposed, i.e. if $\delta=0$ as in standard SFMs, then Eq.\Ref{small} would imply strict \emph{flatness} $\eps_f=0$. This strict flatness has been proven to be one of the main obstacles for recovering classical gravity from SFMs \cite{frankflat,Perini:2012nd,flatness,Han:2017xwo}. But if $\delta\ll1$ is non-zero as above, then small excitations of $\eps_f$ are allowed, and therefore arbitrary smooth curved geometries may emerge from refining triangulations while $\delta\to0$.

It is interesting to note that the opposite limit $\delta\to\infty$ reduce Eq.\Ref{Rcorrrection} to the quantum Regge calculus. Therefore $\delta\in[0,\infty)$ is essentially a parameter interpolating from SFMs to quantum Regge calculus.

The above discusses only the integral with $k_f=m_\t=0$ in $Z_\delta(\ck)$. Nonzero $k_f=m_\t$ shifts $F_f=i\g\eps_f$ in the above computation by $F_f\to F_f+4\pi i \frac{k_f}{q}+\pi i \sum_{\pm}m_\t^\pm \frac{q\pm p}{q}$. In particular Eq.\Ref{small} becomes
\be
\lt|\g{\eps}_f+4\pi \frac{k_f}{q}+\pi  \sum_{\pm}m_\t^\pm \frac{q\pm p}{q}\rt|\lesssim\delta^{1/2}. \label{smallS0}
\ee 
The deficit angles $\eps_f$ are all small for small perturbations $(\mathring{J}+\delta J,\mathring{X}+\delta X)$ of the flat geometry. Therefore, for finite $\g$, Eq.\Ref{smallS0} can only be satisfied for $k_f=m_\t=0$. So integrals in $Z_\delta(\ck)$ with nonzero $k_f=m_\t$ are all suppressed in the perturbative regime unless $\vec{k}=0$.

We would like to remark that the above perturbative study of SFM on a large triangulation by $(\mathring{J}+\delta J,\mathring{X}+\delta X)$ follows from the standard technique in perturbative quantum field theory, i.e. fixing a solution of equation of motion as the background vacuum and perturbing all field variables\footnote{The background $(\mathring{J},\mathring{X})$ satisfies the equations of motion \Ref{eom} and \Ref{small}.}. Our method is different from the boundary state formalism used in e.g. \cite{Bianchi:2010mw,propagator3,propagator2,propagator1,propagator} in the context of a single 4-simplex.

\section{Semiclassical continuum limit:}

\subsection{The idea}

The above discussion is based on a fixed triangulation $\ck$ adapted to a hypercubic lattice. From this, we may construct a refined triangulation $\ck'$ by subdividing each hypercube into 16 identical hypercubes, triangulated by simplices in the same manner as above. By refining the hypercubic lattice, we define a sequence of triangulations $\ck_{N}$ where the label $N$ is the total number of vertices in $\ck$. The continuum limit is $N\to \infty$ in which the vertices in the triangulation become dense in a region of $\R^4$.

We can now associate a SFM $Z_\delta(\ck_N)$ to each $\ck_N$, with $N\to\infty$ as the continuum limit of SFM. The above large spin analysis can be applied to all $Z_\delta(\ck_N)$. This gives a sequence of EOMs \Ref{eom} (or its linearization Eq.\Ref{lin}) and \Ref{small}. All quantities in the equations, e.g. the spins $J_f=J_f(N)$, the regulator $\delta=\delta(N)$, the simplicial geometries, etc, depend on $N$, and flow with $N\to \infty$, which defines the semiclassical continuum limit (SCL). In particular, we will show below that the solutions to the EOM \Ref{eom} flow to solutions of smooth Einstein equation as $N\to \infty$. This can be derived from the fact that the solutions of linearized Regge equation converge to solutions of linearized Einstein equation as the lattice spacing $a\to0$ (see \cite{0264-9381-5-12-007,0264-9381-5-9-004,Dittrich:2007wm}). The EOMs \Ref{eom} are already Regge equation and it only remains to relate the Regge limit $a\to0$ and the SFM continuum limit $N\to \infty$. In fact relating the limits is nontrivial and specifies the SCL. 



The regulator $\delta(N)$ should go to zero with $N\to \infty$ in order to guarantee that the continuum result does not depend on $\delta$. Yet, \Ref{regime} must still be satisfied at every step $N$ for the above asymptotic analysis of $Z_{\delta}(\ck_N)$ to remain valid. Thus, $\l(N)$ has to grow faster than $\delta(N)^{-2}$ (see Subsection \ref{EOLT}).


Recall that the area from SFM is given by $\mathbf{a}_f= \g \l j_f\ell_P^2$. The lattice spacing $a$ relates to background $(\mathring{J}(N),\mathring{X}(N))$ by (recall Section \ref{secSFMs})
\be
a(N)=(\g\l(N))^{1/2}\ell_P.\label{a0000}
\ee
where $\l(N)$ is the typical value of $\mathring{J}_f(N)$ over $\ck$. Note that the background data $(\mathring{J}(N),\mathring{X}(N))$ depends on $\ck_N$, thus depends on $N$.  

We would like to relate Regge continuum limit, so we must require $a(N)\to 0$ as $N\to\infty$. It is possible even when we have $\l(N)\to\infty$ as $N\to\infty$, because we take at the same time the semiclassical limit $\ell_P\to 0$. Practically we define a scaling parameter $\mu(N)$ depending on $N$, and replace $\ell_P$ by
\be
\ell_P\to \mu(N)\ell_P,\  \text{such that}\  \mu(N)\to 0\ \text{as}\ N\to\infty.
\ee
By scaling $\ell_P$, the lattice spacing $a(N)$ in Eq.\Ref{a0000} becomes
\be
a(N)=(\g\l(N))^{1/2}\mu(N)\ell_P.\label{amu}
\ee
The scaling $\mu(N)$ may be viewed as a change of length unit (from small to large), such that the numerical value of $\ell_P$ becomes $\mu(N)\ell_P$ in the new unit. We zooms out to a coarser length unit at the same time as refining the lattice $N\to\infty$, so effectively scales $\ell_P\to 0$. Thus $\mu(N)\to 0$ is understood as an infrared (IR) limit.

To clarify the motivation, it may be illustrating to look at the Regge action term in Eq.\Ref{Rcorrrection} by writing $\mathbf{a}_f(N)=\g J_f\lt(\mu(N)\ell_P\rt)^2$
\be
\lag \g \vec{J}(\ell),\vec{\eps}(\ell)\rag(N)=\frac{1}{\mu^2(N)\ell_P^2}\sum_{f} \mathbf{a}_f(N)\,\eps_f(N)\label{illus}
\ee
$\mu(N)\to0$ as $N\to\infty$ implements both the semiclassical limit in the path integral and the continuum limit of the Regge action. Indeed, if the lattice spacing $a(N)$ satisfies
\be
\lim_{N\to\infty}a(N)= 0,\label{alimit}
\ee 
Eq.\Ref{illus} gives \cite{CMS}
\be
\lag \g \vec{J}(\ell),\vec{\eps}(\ell)\rag(N)= \frac{1}{\mu^2(N)\ell_P^2}\int \rmd^4 x\sqrt{ g}\, R[g]\,(1+\eps(N))
\ee
where $\eps(N)\to 0$ as $N\to\infty$.

Because $\mu(N)$ is a monotonically decreasing function of $N$, we may invert this function and write $N(\mu)$ and make the change of variable to all quantities: 
\be
&&\ck_N=\ck_\mu,\ J(N)=J(\mu),\ X(N)=X(\mu),\nonumber\\ 
&&\l(N)=\l(\mu),\ \delta(N)=\delta(\mu),\ a(N)=a(\mu).
\ee
All previously $N$-dependent quantities becomes $\mu$-dependent. The continuum limit $N\to\infty$ becomes $\mu\to 0$, and Eq.\Ref{alimit} becomes 
\be
\lim_{\mu\to 0} a(\mu)=0.\label{alimit1}
\ee

As a key requirement to relate SFM and Regge continuum limits, Eq.\Ref{alimit1} requires $\rmd a(\mu)/\rmd \mu >0$, which together with \Ref{amu} requires:
\be
-\frac{2}{\mu}<\frac{1}{\l}\frac{\rmd \l}{\rmd\mu}<0.\label{ineq1}
\ee

The inequality \Ref{ineq1} is not the only requirement in order to relate to the Regge continuum limit. Recall that solutions of Regge equation arise in the leading order stationary phase approximation of $Z(\ck_\mu)$ as $\l(\mu) \gg1$. The solutions have the (quantum) corrections of $O(1/\l)$. The correction is bounded by $C(\mu)/\l(\mu)$ with $C(\mu)>0$, where $C(\mu)$ grows as $\mu\to0$ (see Section \ref{EOLT}). As a result, $\l(\mu)$ is required to grow in a faster rate, in order to keep $C(\mu)/\l(\mu)$ small to suppress the $1/\l$ correction to Regge solutions as $\mu\to0$. It implies
\be
\frac{1}{\l}\frac{\rmd \l}{\rmd\mu}<\frac{1}{C}\frac{\rmd C }{\rmd\mu}.\label{ineq2}
\ee


In addition to the constraints \Ref{ineq1} and \Ref{ineq2}, it follows from \Ref{small} and $\eps_f\simeq a^2/\rho^2$ that there should exist a bound $L<\infty$ s.t. 
\be
\frac{\delta(\mu)^{1/2}}{a(\mu)^2}< L\label{ineq3}.
\ee 
Otherwise, the curvature of the emergent geometry (i.e. $\rho^{-2}= \lim\eps_f(\mu)/a(\mu)^2$) would diverge. \\

\begin{Definition} 
A semiclassical continuum limit (SCL) is the flow of the 3 parameters $\l(\mu),a(\mu)$ and $\delta(\mu)$ as $\mu\to 0$ (together with the lattice refinement) that satisfy \Ref{ineq1}, \Ref{ineq2}, and \Ref{ineq3}. $a(\mu)$ and $\delta(\mu)$ tend to zero in the limit $\mu\to0$, while $\l(\mu)\to\infty$ grows faster than $\delta(\mu)^{-2}$. 
\end{Definition}

The SCL is well-defined although \Ref{ineq1}, \Ref{ineq2}, and \Ref{ineq3} indeed give nontrivial restrictions.

\begin{Theorem}

The SCL is well-defined because the flows satisfying the requirements always exist. 

\end{Theorem}

The proof of the above statement is given in Subsection \ref{running}. An SCL relates the SFM continuum limit to the Regge continuum limit, and allows us to apply the convergence in Regge calculus to geometries coming from SFM critical points.

\subsection{Expansion of the linearized theory}\label{EOLT}

The large spin analysis uses the stationary phase approximation, which is an $1/\l$ asymptotic expansion of integrals in $Z_\delta(\ck)$. We focus on the expansion of the integral with $k_f=m_\t=0$, at the level of the linearized theory. 

We write $\delta X=\delta X(\ell)+\delta\cx$, where $\delta X(\ell)$ solves the critical equations $\delta_X S=\mathrm{Re}(S)=0$. By this change of variables, 
\be
S=S\lt[\mathring{\ell}+\delta\ell,\mathring{X}+\delta X(\ell)\rt]+\half\delta \cx^T H_{\cx\cx}\delta \cx +\cdots.\label{quadratic0}
\ee

From the discussion in the last section, we know that $S[\mathring{\ell}+\delta\ell,\mathring{X}+\delta X(\ell)]$ is the Regge action. At the quadratic order,
\be
S[\mathring{\ell}+\delta\ell,\mathring{X}+\delta X(\ell)]=\half\delta\ell^T H_{\ell\ell}\delta\ell+\cdots
\ee  
has been studied in \cite{Rocek:1982tj}, in which the Hessian matrix $H_{\ell\ell}$ was shown to be degenerate. The kernel of the Hessian contains (1) the space of solutions of linearized Regge equation, and (2) 4 zero modes corresponding to the diffeomorphisms in the continuum, and (3) 1 zero mode of hyperdiagonal edge-length fluctuation. 



We obtain the following bound of error for the large spin analysis in the last section \footnote{Eq.\Ref{bound1} assumes the non-degeneracy of the Hessian matrix $H_{\cx\cx}$ after removing gauge redundancies. This non-degeneracy is supported by some numerical evidences. A general proof of non-degeneracy for the Hessian in SFM is lacking in the literature. In case it happens that $H_{\cx\cx}$ is degenerate, there are additional zero modes coming from SFM variables $\cx$. Then the effective theory is the Regge gravity coupling to these additional zero modes, when go beyond the linearization. But in this paper, we focus on the sector of linearized Regge gravity and the continuum limit. }
\begin{widetext}
\be
\lt|\int[\rmd\ell\rmd X]\, e^{\l S}D_{\delta}(\ell,X)-\lt(\frac{2\pi}{\l}\rt)^{\frac{\cn}{2}}\lt[\det\lt(H_{\cx\cx}\rt)\det \lt(K_{\ell\ell}^\perp\rt)\rt]^{-\half}\int[\rmd \delta \ell^\parallel]\, D_{\delta}\lt(\delta\ell^\parallel,X(\delta\ell^\parallel)\rt) \rt|\leq \lt(\frac{2\pi}{\l}\rt)^{\frac{\cn}{2}}\frac{C}{\l}.\label{bound1}
\ee
\end{widetext}
Here $K_{\ell\ell}^\perp$ is the nondegenerate part of $H_{\ell\ell}$, and $\cn=\rank(K_{\ell\ell}^\perp)+\rank(H_{\cx\cx})$. The integral $\int[\rmd \delta \ell^\parallel]$ is over solutions of linearized Regge equations and zero modes. $C>0$ bounds the $1/\l$ correction \cite{stationaryphase}. The semiclassical approximation by Regge solutions is valid when the $1/\l$ corrections are negligible, i.e. when ${C}^{\delta}/{\l}$ is small.

The bound relates to the derivatives of $D_\delta$ by \cite{stationaryphase}
\be
\frac{C}{\l}=\frac{c}{\l}\lt(\mathrm{sup}\,\big|\partial D_\delta\big|+\mathrm{sup}\,\lt|\partial^2 D_\delta\rt|\rt).
\ee 
where $c$ is a constant. Since $\partial^2 D_\delta\sim \delta^{-2}$, 
\be
\l\gg\delta^{-2}\gg1\label{regimeS}
\ee 
has to be satisfied to validate the expansion.


Eq.\Ref{bound1} is the expansion at the level of linearized theory, whose asymptotics is an integral over critical solutions (solutions of EOM and zero modes). It indicates that the critical solutions contribute dominantly to the SFM. In this paper we mainly discuss the convergence of critical solutions under the semiclassical continuum limit. In a companion paper \cite{HHZprop}, we report the result of graviton propagator and the continuum limit, in which we apply gauge fixings to remove zero modes.

\subsection{Semiclassical Continuum Limit (SCL)}\label{running}

We construct a refined triangulation $\ck'$ which is adapted to a refined hypercubic lattice in the same way as $\ck$ is adapted to the original hypercubic lattice. The refined hypercubic lattice is given by subdividing each hypercube into 16 identical hypercubes. By refining the hypercubic lattice we define a sequence of triangulations $\ck_\mu$ where $\ck_{\mu'}$ is finer than $\ck_\mu$ if $\mu'<\mu$. In the continuum limit $\mu\to 0$  the vertices in the triangulation become dense in $\R^4$. 

A sequence of SFMs is defined by associating an amplitude $Z(\ck_\mu)$ to each $\ck_\mu$. Since the above large spin analysis is valid for all $Z(\ck_\mu)$, it gives a sequence of Eqs.\Ref{eom} and \Ref{small} on the sequence of $\ck_\mu$:
\be
\sum_f\frac{\partial\mathbf{a}_f(\mu)}{\partial \ell} \eps_f(\mu)=0,\quad |\g{\eps}_f(\mu)|\lesssim\delta^{1/2}(\mu).\label{eommu}
\ee
All quantities in the equations, e.g. the spins $J_f$, the regulator $\delta$, and the simplicial geometries, etc, depend on $\mu$, and flow toward $\mu\to 0$. 

We set the triangulation label $\mu$ to be a mass scale such that $\mu^{-1}$ is a new length unit. Then $\mathbf{a}_f(\mu)=\a_f(\mu)\mu^{-2}$. The lattice spacing $a(\mu)$ is given by the background flat geometry on $\ck_\mu$:
\be
\mathring{\ell}(\mu)=\lt(\g\l(\mu)\rt)^{\half}\ell_P=a(\mu)\mu^{-1}.\label{amuS}
\ee

We define the semiclassical continuum limit (SCL) as the flow of the 3 parameters $\l(\mu),a(\mu),\delta(\mu)$, where $a(\mu),\delta(\mu)\to0$ and $\l(\mu)\to\infty$ ($\l(\mu)\gg\delta(\mu)^{-2}$) for $\mu\to0$. In addition, these flows should satisfy
\be
&&-\frac{2}{\mu}<\frac{1}{\l}\frac{\rmd \l}{\rmd\mu}<0,\label{ineq1S} \\
&&\frac{1}{\l}\frac{\rmd \l}{\rmd\mu}<\frac{1}{C}\frac{\rmd C}{\rmd\mu},\label{ineq2S}\\
&&\frac{\delta(\mu)^{1/2}}{a(\mu)^2}\ \text{bounded from above}.\label{ineq3S}
\ee
Here, $C(\mu)$ is the bound in Eq.\Ref{bound1}, which now depends on $\mu$ for the expansion of $Z(\ck_\mu)$.

The constraint Eqs.\Ref{ineq1S} - \Ref{ineq3S} are necessary due to the following reasons: Firstly, the motivation for the SCL is to relate the SFM continuum limit $\mu\to0$ to the continuum limit $a\to0$ in Regge calculus, so that we can apply the convergence result in Regge calculus to the solutions of Eqs.\Ref{eom} and \Ref{small}. Obviously, this requires that the lattice space $a(\mu)^2\propto\l(\mu)\mu^2\to0$ as $\mu\to0$. Thus,
\be
0<\frac{\rmd}{\rmd\mu}\lt(\l(\mu)\mu^2\rt)=\mu^2\frac{\rmd\l}{\rmd\mu}+2\mu \l
\ee
which yields Eq.\Ref{ineq1S}.

Secondly, the $1/\l$ correction has to be small for all $\mu$, in order that classical Regge solutions are the leading orders of $Z(\ck_\mu)$. It is important to have Regge solutions at all $\mu$ to apply the convergence result in Regge calculus. This demands Eq.\Ref{bound1} to be valid for all $Z(\ck_\mu)$ with $C(\mu)/\l(\mu)$ being always small. 

$C(\mu)\sim \delta(\mu)^{-2}$ grows when the triangulation is refined. Thus, $\l(\mu)$ is required to grow in a faster rate in order to suppress $C(\mu)/\l(\mu)$ as $\mu\to0$. This requires
\be
0<\frac{\rmd}{\rmd\mu}\lt(\frac{C(\mu)}{\l(\mu)}\rt)=-\frac{C}{\l^2}\frac{\rmd \l}{\rmd\mu}+\frac{1}{\l}\frac{\rmd C }{\rmd\mu}
\ee
which gives
\be
\frac{1}{\l}\frac{\rmd \l}{\rmd\mu}<\frac{1}{C}\frac{\rmd C }{\rmd\mu}.\label{app1}
\ee
This condition guarantee that Eq.\Ref{bound1} is valid at all $\mu$, with the $1/\l$ correction being always small, i.e. the following bound holds in the continuum limit $\mu\to0$:
\be
\frac{C(\mu)}{\l(\mu)}<\frac{C(1)}{\l(1)},\label{smallness}
\ee
where $\mu=1$ is the starting point of the flow.

Thirdly,  the simplicial geometry should approximates a smooth geometry. If this is the case then the typical curvature radius $\rho$ of the smooth geometry relates to the deficit angle of the simplicial geometry by $\rho^{-2}\simeq\eps_f  a^{-2}$. The regulator $\delta$ and conditions \Ref{small} and \Ref{ineq3S} guarantee that the curvature $\rho^{-2}$ of the emergent geometry is bounded (geometry is nonsingular) as $\mu\to0$.

Eqs.\Ref{ineq1S} - \Ref{ineq3S} have nontrivial implications for the SCL: In order that a satisfactory flow $\l(\mu)$ exists, Eqs.\Ref{ineq1S} and \Ref{ineq2S} have to be consistent, i.e. 
\be\label{app2}
\frac{1}{C}\frac{\rmd C }{\rmd\mu}> -\frac{2}{\mu}, \label{consistencyS}
\ee
which yields a restriction to the assignment of $\mu$ to $\ck_\mu$. Since  $\mu$ is assigned to a sequence of triangulations $\ck_{\mu}\equiv\ck_{\mu_n}\equiv\ck_{n}$ ($\mu_{n-1}>\mu_{n}$), the variable $\mu\equiv\mu_n$ is actually discrete. In the above, we have assumed that $C(\mu_n)$ and $\l(\mu_n)$ can be continued to differentiable functions $C(\mu)$ and $\l(\mu)$.
 Integrating Eq.\Ref{app2} leads to
\be
\int^{\mu_{n-1}}_{\mu_n}\frac{1}{C}\frac{\rmd C }{\rmd\mu}\rmd\mu> -\int^{\mu_{n-1}}_{\mu_n}\frac{2}{\mu}\rmd\mu
\ee
which implies the following constraint on $\mu_n$:
\be
\frac{\mu_{n-1}}{\mu_n}>\lt[\frac{C(\mu_n)}{C(\mu_{n-1})}\rt]^{\frac{1}{2}}.\label{app3}
\ee
Note that, $\mu_n$ satisfying this constraint always exists. 

Once we have a satisfactory assignment of $\mu$ to $\ck_\mu$, the running behavior of $\l(\mu)$ is constrained by
\be
-\frac{2}{\mu}<\frac{1}{\l}\frac{\rmd \l}{\rmd\mu}<\frac{1}{C}\frac{\rmd C}{\rmd\mu}.\label{cons1S}
\ee
In addition, Eqs.\Ref{ineq3S} and \Ref{regimeS} requires $\delta(\mu)$ to satisfy
\be
\l(\mu)^{-1/2}\ll \delta(\mu)\leq L^2 \l(\mu)^2\mu^4
\ee
where $L\g^{-2}\ell_P^{-2}$ is the bound of ${\delta(\mu)^{1/2}}/{a(\mu)^2}$. The existence of a satisfactory $\delta(\mu)$ requires that 
\be
\l(\mu)^{5/2}\gg\mu^{-4}.\label{cons2S}
\ee
which is another constraint for the flow $\l(\mu)$.

A flow $\l(\mu)$ satisfying both constraints Eqs.\Ref{cons1S} and \Ref{cons2S} always exists. The following provides a satisfactory example of $\l(\mu)$. Consider the ansatz: 
\be
\l(\mu)=\l(1)\, \mu^{-2+u},\label{ansatz}
\ee
where $\l(1)$ is the initial value of $\l(\mu)$ at $\mu=1$. Eq.\Ref{cons1S} implies
\be
u>0,\quad \frac{1}{ C}\frac{\rmd C }{\rmd\mu}> -\frac{2-u}{\mu},\quad\forall 1\leq s\leq m+1.
\ee
The second inequality certainly can be satisfied by a suitable assignment of $\mu$ to $\ck_\mu$, by a similar derivation showing Eq.\Ref{consistencyS} can be satisfied (replacing $\frac{2}{\mu}$ by $\frac{2-u}{\mu}$). It doesn't restrict the value of $u$. But combining \Ref{cons2S}, we obtain an upper bound of $u$:
\be
0<u<\frac{2}{5}.
\ee
If $u$ is within the above range then we obtain a satisfactory flow $\l(\mu)=\l(1) \mu^{-2+u}$, which implies $a(\mu)=\mu^{u/2}\sqrt{\g \l(1)\ell_P^2} $, and $\l(1)^{-1/2}\mu^{1-u/2}\ll \delta(\mu)\leq L^2 \mu^{2u}$. This example illustrates that flows $\l(\mu),a(\mu),\delta(\mu)$, which satisfy Eqs.\Ref{ineq1S} - \Ref{ineq3S} always exist. So the SCL of SFM is well-defined. 



\section{Emergent Linearized gravity:} The above SCL fills the gap between the continuum limits in SFM and Regge calculus. Thus, the sequence of critical points satisfying Eq.\Ref{eom} under the SCL is the same as the sequence of Regge solutions under $a\to0$.

The classification of Linearized Regge solutions and their convergence has been studied in \cite{0264-9381-5-12-007,0264-9381-5-9-004} (reviewed in Appendix \ref{sec:GravityWave}). It is shown that the solutions of linearized Regge equation converge to smooth solution of 4d (Riemannian) Einstein equation in the limit $a\to 0$. All the nontrivial geometries obtained from the limit have curvatures as linear combinations of 
\be
R_{abcd}(x)=\mathrm{Re}\lt[W_{abcd} \exp\lt(-k\cdot x\rt)\rt],\label{GW}
\ee 
which are Euclidian analogs of plane waves. Here $k\cdot x$ is the 4d Euclidean inner product and $k\in\C^4$ satisfy  $k\cdot k=0$. $W_{abcd}$ is a traceless constant tensor that spans a 2-dimensional solution space, whose dimensions correspond to the helicity $\pm2$ gravitons.

Recall that the main contributions to $Z(\ck_{\mu})$ in the SCL come from critical points that satisfy linearized Regge equation, all other contributions are suppressed. Moreover, the SCL maps the SFM IR limit $\mu\to 0$ to Regge calculus limit $a\to 0$. Therefore, the above convergence result of Regge solutions can be applied to SFM as $\mu\to 0$, which shows that on a 4d flat background, the low energy excitations of SFM give all smooth solutions of linearized Einstein equation (gravitons).

\section{Conclusion and Outlook:} In the above discussion, we have shown that from the SCL, the low energy excitations of SFM on a flat background give all smooth (linearized) Einstein solutions. It indicates that linearized classical gravity is the effective theory emerging from SFMs at low energy. Our result indicates that the SFM, being a discrete model of fundamentally entangled qudits, is a working example for the idea in emergent gravity program.   

Here we showed for the first time that smooth curved spacetimes can emerge from SFMs in a suited continuum limit. 
It suggests that SFMs have a proper semiclassical limit not only at the discrete level but also in the continuum. Our result, therefore, strengthens the confidence that covariant LQG is a consistent theory of quantum gravity.

As a key technical tool, a regularization/deformation of the SFM is employed in the derivation. This deformation interpolates between SFMs and quantum Regge calculus, and the deformation parameter $\delta$ becomes a coupling constant of a higher curvature correction term to Einstein gravity from SFM.

Our analysis certainly can be generalized to the nonlinear regime, and even to the case of strong gravitational field. Indeed the large spin analysis doesn't rely on the linearization, and the EOM \Ref{eom} is nonlinear. The emergence of black hole or cosmological solutions from SFMs can be derived by applying the Regge calculus convergence results in e.g. \cite{Gentle:2002ux}, similarly as above. These solutions will enable us to study singularities as the high energy excitations from SFMs. 

Finally we remark that the flows of SFM parameters $\l(\mu),a(\mu),\delta(\mu)$ in the SCL likely relate to a renormalization group flow\footnote{It may relate to the recent development of the renormalization group flow in SFM \cite{Bahr:2017klw}.}. Further investigation of the relation may shed light on the renormalization of perturbative gravity.

\section*{Acknowledgements}

MH acknowledges Abhay Ashtekar, Jonathan Engle, Ling-Yan Hung, Simone Speziale, and Qiang Wen for the helpful discussions and comments. AZ acknowledges support through the Alexander von Humboldt foundation through a Feodor Lynen stipend. This work receives support from the US National Science Foundation through grant PHY-1602867, and Start-up Grant at Florida Atlantic University, USA.

\appendix

\section{Spin Foam Models (SFMs) and Tensor Networks}\label{SFMTN}

\begin{figure}[h]
	\centering\includegraphics[width=0.8\linewidth]{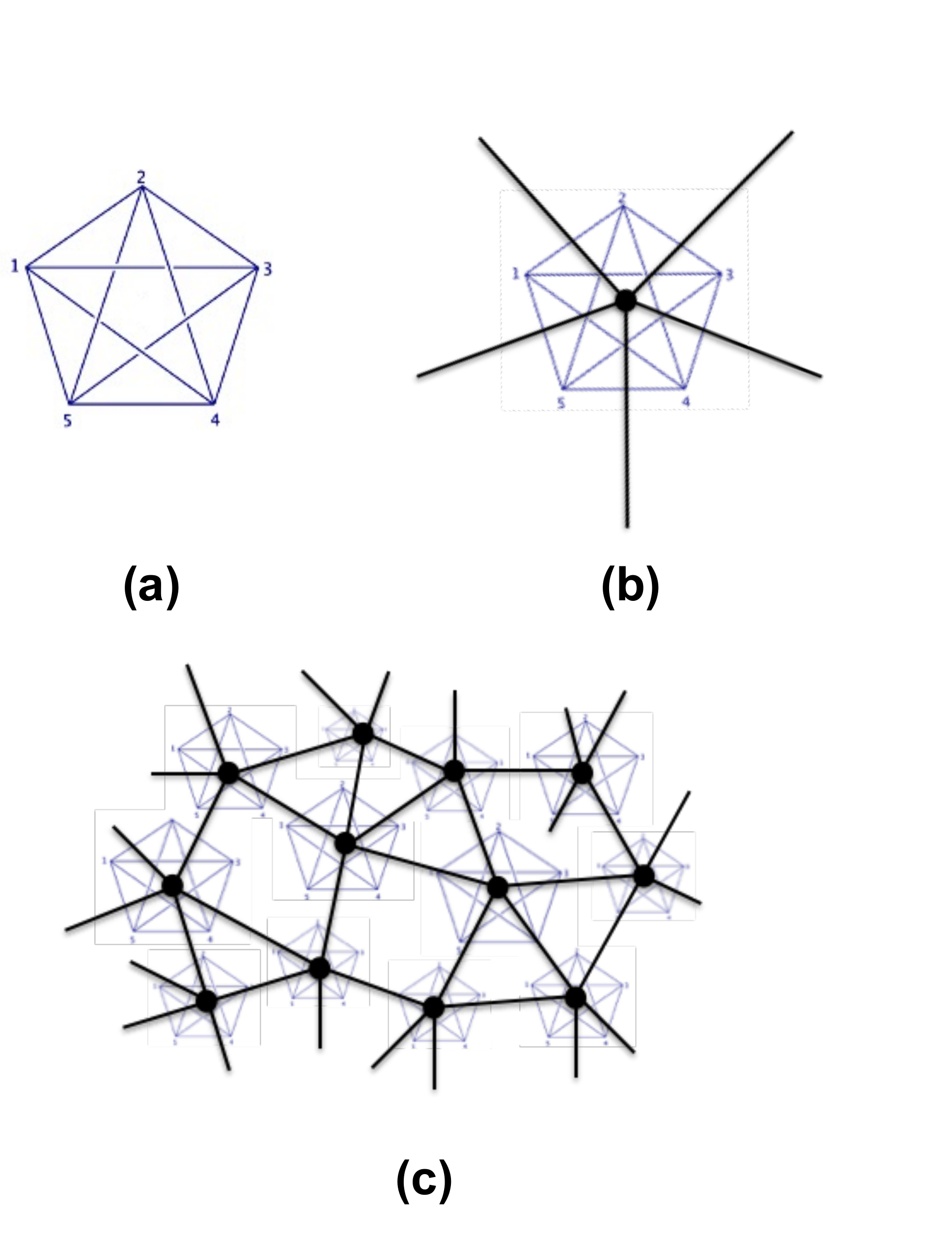}
	\caption{(a) A 4-simplex $\sig$ as the building block of 4d triangulation $\ck$. (b) The 5-valent vertex illustrates a rank-5 tensor $|A_\sig(\vec{J})\rangle$. (c) Gluing 4-simplices $\sig$ in $\ck$ gives a tensor network $\mathrm{TN}(\ck,\vec{J})$.}
	\label{simplex}
\end{figure}

In 4 dimensions, the main building block of a triangulation $\ck$ is a 4-simplex $\sig$ (see FIG.\ref{simplex}(a)), whose boundary  $\partial\sig$ contains 5 tetrahedra $\t$ and 10 triangles $f$.  $\ck$ is obtained by gluing a (large) number of $\sig$ through pairs of their boundary tetrahedra. 
In the following $\ck$ itself should be understood as purely combinatorial or topological while the discrete geometry is encoded in the associated state sum $Z(\ck)$ of the SFM. 
%
Generically, $Z(\ck)$ takes the form 
\be
Z(\ck)=\sum_{\vec{J}}\sum_{\vec{i}}\prod_fA_f(J_f)\prod_\sig A_\sig(J_f,i_\t), \label{ZK}
\ee
where the summand products over all triangles $f$ and all 4-simplices $\sig$ in the triangulation $\ck$. The SFM data $(\vec{J},\vec{i})$ assigns each triangle $f$ an SU(2)-representation labelled by $J_f\in \Z_+/2$, and assigns each tetrahedron $\t\subset\ck$ an SU(2)-intertwiner (rank-4 invariant tensor) $i_\t$, i.e. 
\be
i_\t \in \inv{V_{J_1}\otimes\cdots\otimes V_{J_4}}\equiv\ch^{inv}_{J_1\cdots J_4} .
\ee 
Each $\sig$ associates to a \emph{4-simplex amplitude} $A_\sig(J_f,i_\t)$, which depends on 10 $J_f$ and 5 $i_\t$ assigned to $f,\t \subset\partial\sig$.  The weight $A_f(J_f)$ of $\sum_{\vec{J}}$ is usually called the \emph{face amplitude}. 


Both, the face amplitude $A_f(J_f)$ and the 4-simplex amplitude $A_\sig(J_f,i_\t)$ are model dependent. In the following we mainly focus on the Euclidean Engle-Pereira-Rovelli-Levine (EPRL) model \cite{EPRL,FK}.  In this model, the 4-simplex amplitude $A_{\sig}$ is given by the contraction of 5  $\mathrm{Spin}(4)$ invariant tensors $I_{\t}$, that depend on  $i_\t$ ($\t=1,\cdots 5$). That is, 
\be
A_{\sig}(J_f,i_\t)=\tr\lt(I_{1}\otimes I_{2}\otimes I_{3}\otimes I_{4}\otimes I_{5}\rt),
\ee
where $I_\t$ is given by
\be
I^{m_1^\pm\cdots m_4^\pm}_\t=\int\rmd h^+\rmd h^- \prod_{f=1}^4 \lt[D^{(J^+_f)}_{m^+_fn^+_f}(h^+)D^{(J^-_f)}_{m^-_fn^-_f}(h^-)C^{n_f^+n_f^-}_{n_f}\rt]i_\t^{n_1\cdots n_4}\nonumber.
\ee
The above integral integrates over 2 copies of SU(2) with Haar measures $\rmd h^{\pm}$. $D^{(J^\pm_f)}_{m^\pm_fn^\pm_f}(h^\pm)$ are Wigner $D$-matrices for the representation $J^\pm_f$ and $C^{n_f^+n_f^-}_{n_f}$ are Clebsch-Gordan coefficients interpolating between $(J^+_f,J^-_f)$ and $J_f$ ($f=1,\cdots,4$) which are subject to the constraint 
\be
J^{\pm}_f=\half|1\pm\gamma|J_f.
\ee
Here, $\g\in\R$ is the Barbero-Immirzi parameter. If $\g=p/q$ $(p,q\in\Z)$, then $J^{\pm}_f\in\Z/2$ implies $J_f\in q\Z$ for $p+q$ odd or $J_f\in q\Z/2$ for $p+q$ even.

Note that $A_\sig(J_f)$ with fixed $J_f$'s  is a linear map from 5 invariant tensors $i_\t$ to $\C$. In other words, $A_\sig(J_f)$ is a rank-5 tensor state (see FIG.\ref{simplex}(b))
\be
&&|A_\sig(J_f)\rangle \in \nonumber\\
&&\ch^{inv}_{J_1J_2J_3J_4}\otimes \ch^{inv}_{J_4J_5J_6J_7}\otimes \ch^{inv}_{J_7J_3J_8J_9}\otimes \ch^{inv}_{J_9J_6J_2J_{10}}\otimes \ch^{inv}_{J_{10}J_8J_5J_1}.
\ee
Thus, the 4-simplex amplitude can be written as an inner product
\be
A_{\sig}(J_f,i_\t)=\langle i_1\otimes\cdots\otimes i_5|A_\sig(J_f)\rangle.
\ee

The above relation allows us to write the summand of $\sum_{\vec{J}}$ in Eq.\Ref{ZK} as a tensor network. We observe that  a pair of $\sig,\sig'$ is glued in $\ck$ by identifying a pair of tetrahedra $\t=\t'=\sig\cap\sig'$. Correspondingly, a pair of invariant tensors in $A_\sig,A_{\sig'}$ is identified by setting $i_\t=i_{\t'}$ and summing over  $i_\t$ in  $Z(\ck)$. The identification and summation may be formulated by inserting a maximal entangled state at each $\t$
\be
|\t\rangle:=\sum_{i_\t}|i_\t\rangle\otimes|i_\t\rangle\in \ch^{inv}_{J_1J_2J_3J_4}\otimes\ch^{inv}_{J_1J_2J_3J_4}.
\ee
This yields  a tensor network, 
\be
\mathrm{TN}(\ck,\vec{J}):=\otimes_\t\langle\t| \otimes_\sig|A_\sig(J_f)\rangle,\label{TN}
\ee
where the tensors $A_\sig$  at the vertex is contracted  with $|\t\rangle$ at the edges (see FIG.\ref{simplex}(c)).
In other words, the EPRL pair of $|i_\t\rangle$ in $|\t\rangle$ is associated to the two ends of the edge in FIG.\ref{simplex}(c), and contracted with the pair $|A_\sig\rangle,|A_{\sig'}\rangle$ ($\sig\cap\sig'=\t$) at the two ends \footnote{To compare with the usual definition of SFM, the network in FIG.\ref{simplex}(c) is the 1-skeleton of the 2-complex dual to $\ck$. Note that the network FIG.\ref{simplex}(c) was oriented in the usual definition of SFM, where $i_\t$ associated to the target of each edge was the dual $\langle i_\t|$. Here we have encoded the duality map $|i_\t\rangle\mapsto\langle i_\t|$ at the target of each oriented edge into $|A_\sig(\vec{J})\rangle$, in order to formulate $\mathrm{TN}(\ck,\vec{J})$ as a projected entangled pair states (PEPS) \cite{Verstraete:2004cf,Qi1}. }. 
Inserting \Ref{TN} into \Ref{ZK} finally gives
\be
Z(\ck)=\sum_{\vec{J}}\mathrm{TN}(\ck,\vec{J})\prod_fA_f(J_f).\label{ZTN}
\ee
Note that both, $\mathrm{TN}(\ck,\vec{J})$ and $Z(\ck)$, are wave functions of boundary SFM data if $\partial\ck\neq\emptyset$, or numbers if $\partial\ck=\emptyset$.

Due to the presence of the maximal entangled states $|\t\rangle$, the tensor network formulation \Ref{ZTN} allows to interpret SFMs as models of entangled qubits (or more precisely qudits). Recent advances in condense matter suggest that entangled qubits and their quantum information might be fundamental, while gravity might be emergent phenomena (see e.g. \cite{2017arXiv170903824W}). Our results demonstrate that SFMs are concrete examples, in which gravity emerges from fundamentally entangled qubits, and therefore relate quantum gravity to quantum information.

An important step in establishing the results of this paper is to analyze the behavior of \Ref{ZTN} for large spins. This is best studied in the integral representation of $\mathrm{TN}(\ck,\vec{J})$ \cite{semiclassicalEu,HZ1}: 
\be
\mathrm{TN}(\ck,\vec{J})=\int\rmd g^{\pm}_{\sig\t}\rmd \xi_{\t f}\ e^{\sum_f J_f F_f\lt[g^{\pm}_{\sig\t},\, \xi_{\t f}\rt]}
\ee
where $g^{\pm}_{\sig\t}\in\Su\times\Su$ and $\xi_{\t f}\in\C^2$ are normalized spinors,  $<\cdot|\cdot>$ is the Hermitian inner product and $F_f$ is expressed as
\be\label{eq:actiong1}
&&F_f\lt[g^{\pm}_{\sig\t},\, \xi_{\t f}\rt]=\sum_{\sig, f\subset\sig}\Big[(1-\g)\ln\lag{\xi_{\t f}}\big|(g^-_{\sig\t})^{-1}g^-_{\sig\t^\prime}\big|{\xi_{\t^\prime f}}\rag\nonumber\\
	&&\quad+(1+\g) \ln\lag{{\xi_{\t f}}\big|(g^+_{\sig\t})^{-1}g^+_{\sig\t'}\big|{\xi_{\t' f}}}\rag\Big].
\ee
The above integral representation is valid for $\gamma<1$. For $\g>1$ one obtains a similar expression (see  \cite{HZ1}).

\section{Convergence to Smooth Geometry}\label{sec:GravityWave}

The equations of motion from SFM's contain the Regge equation
\be
\sum_f\frac{\partial\a_f(\mu)}{\partial \ell} \eps_f(\mu)=0.
\ee
In the  SCL the lattice spacing $a(\mu)$ goes to zero with $\mu\to0$. Therefore the behavior of SFM critical points in the SCL is closely related to the convergences of solutions to Regge equation in the continuum limit $a\to0$. The latter has been studied in \cite{0264-9381-5-12-007,0264-9381-5-9-004} for the linearized theory on a flat background. 

In the following, we review the results in \cite{0264-9381-5-12-007,0264-9381-5-9-004} and apply them to our case. The following discussion often suppresses the label $\mu$ but uses the lattice space $a$ to label the continuum limit. 



 Regge's equation can be written as a set of linear equations of $\eps_f$ for small perturbations on a flat background, i.e.
\be\label{eq:ReggeEOM}
\sum_{{f, \ell\subset f}}\eps_f\cot\lt(\vth_\ell\rt)=0,
\ee
where $\vth_\ell$ is the internal angle of the triangle $f$ opposite to the edge $\ell$ and evaluated on the flat background  $\mathring{\ell}$. 

In addition to Eq.\Ref{eq:ReggeEOM} the deficit angles $\eps_f$ should satisfy the (linearized) Bianchi identity
\be\label{eq:LinearBianchi}
\sum_{{f, \ell\subset f}}\eps_f \lt[U_{[bc}m_{a]} U^{de}\rt]_f=0
\ee
where $m$ is the outward-normal of $\ell$ in the plane of $f$ and $\uab$ is an antisymmetric tensor associate to $f$ that is given by 
\be
U_{ab}=v_aw_b-v_bw_a.
\ee 
The unit vectors $v,w $ are mutually orthogonal and orthogonal to $f$. 

Note that the Bianchi identity is satisfied automatically, if one uses edge-length variables to describe the system. However, here it is more convenient to use deficit angles as the system variables. In this case, the Bianchi identity is an additional constraint. This formulation of linearized Regge equation using deficit angles is equivalent to the one using edge lengths, because  a set of linearized deficit angles satisfying the Bianchi identities can construct a linearized (piecewise-flat) metric, unique up to linearized diffeomorphisms (4 zero modes mentioned in Section \ref{EOLT}). \cite{BARRETT1987135}. Conversely, from the linearized metric, one can construct the linearized deficit angles.

Given the periodic nature of the triangulation $\ck$, we consider the periodic configuration of $\eps_f$ with the shift $\o_i(a)$ along the body principles of a hypercube. The shift relates $\eps_f$ and $\eps_{f'}$ for parallel $f$ in two neighboring hypercubes by 
\be
\eps_f=\o_i(a)\ \eps_f'. \label{eps}
\ee
Here $i=1,2,4,8$ label the 4 body principles of the hypercube. Eq. \Ref{eps} can be more conveniently written by introducing the short hand notation 
\be
\Omega(a)=\lt(\,\o_1(a),\o_2(a),\o_4(a),\o_8(a)\,\rt).
\ee
%
and computing the Fourier transform of Eq.\Ref{eps} on the hypercubic lattice $(a\Z)^4$. This yields,
\be
\eps_f(n)=\int_{-\frac{\pi}{a}}^{\frac{\pi}{a}}\frac{\rmd^4 k}{(2\pi)^4}\,e^{i\sum_ik_in_i a}\eps_f(k),\quad n_i\in\Z.
\ee
Thus each ``plane wave'' corresponds to $\Omega(a)=\lt(\,e^{ik_1 a},e^{ik_2 a},e^{ik_4 a},e^{ik_8 a}\,\rt)$ and tends to $(1,1,1,1)$ in the limit $a\to0$. In the following we will assume the same limit behavior, i.e. $\Omega(a)\to (1,1,1,1)$, and that the derivative $\Omega'(0)$ exists at zero for general $\O(a)$. Note that $\O(a)$ is complex because the $k_i$ 's are complex in Euclidean signature, as we see in a moment. 

Due to periodicity, Eqs.\Ref{eq:ReggeEOM} and \Ref{eq:LinearBianchi} reduces to a set of linear equations for $\eps_f$'s within a single hypercube. Let this hypercube be denoted  by cell(0), then
\be
\sum_{f\subset\text{cell}(0)}M\lt[\O(a)\rt]_{f'f}{\eps}_{f}(a)=0.\label{MepsS}
\ee
In the following we consider complex solutions of the above equation and their convergence. The physical solutions are the real parts of those solutions, and converge when the complex solutions converge.

By selecting a solution $\eps_f(a)$ of Eq.\Ref{MepsS} for each $a$ we can generate a sequence of linearized Regge configurations. The convergence of this sequence is closely related to the convergence of the associated discrete Riemann curvature tensors. The discrete curvature is defined as a tensor-valued distribution that maps  a smooth function $\ff$ of compact support to the tensor $ R_{abcd}[\ff] $ given by 
\be\label{eq:linearRegge}
\ff\rightarrow \sum_{f} \eps_f [\uab \ucd]_f \int_f \ff\zeta\equiv R_{abcd}[\ff] .
\ee
Here $\zeta$ is the area measure of $f$ and  $\uab$ is the bi-vector of the triangle $f$. One can now show that the sequence of solutions to Eq.\Ref{MepsS} converges for $a\to0$ if $R_{abcd}$ converges as a distribution provided that $\eps_f/a^2$ remains bounded \cite{0264-9381-5-12-007,0264-9381-5-9-004}. Note that in the SCL defined above the latter condition is automatically satisfied due to the regulator $\delta$ and Eq.\Ref{ineq3S}.


 
It is more convenient to consider a stronger convergence for the sequence of solutions  ${\eps}_f(a)$. Namely we require that ${\eps}_f(a)/a^2$ converges for all $f$ as $a\to0$, which clearly implies the above convergence criterion.

 In  \cite{0264-9381-5-9-004} it was shown that for any family  of vectors $\O(a)$, for which $\O(0)=(1,1,1,1)$ and  $\O'(0)$ exist, and any
 solution ${\eps}_f^{(0)}$ of Eq.\Ref{MepsS} at a finite $a_0$  there exists a sequence of solutions ${\eps}_f(a)$ of Eq.\Ref{MepsS} such that ${\eps}_f(a_0)={\eps}_f^{(0)}$. Moreover, the limit ${\eps}_f(a)/a^2$ as $a\to0$ exists for all $f$ and the discrete curvature tensor $R_{abcd}$ converges to
\be
R_{abcd}(x)\to W_{abcd} \exp\lt(-\O'(0)\cdot x\rt),\label{GWS}
\ee 
where $W_{abcd}$ is a traceless complex constant tensor, and $\cdot$ is the 4d Euclidean inner product.

There are 3 possible cases for different $k\equiv\O'(0)\in\C^4$. Case 1: If $k\neq0$ satisfies $k\cdot k=0$ then  $W_{abcd}$ spans a 2-dimensional solution space, where the dimension corresponds to the helicity $\pm2$ of gravitons. Note that $k$ has to be complex, otherwise $k\cdot k=0$ would imply $k=0$.

Let $U$ and $V$ denote the real and imaginary part of the tensor $W$, and $m$ and $l$ the real and imaginary part of $k$. The real part of Eq.\Ref{GWS} is
\be\label{eq:real}
&&U_{abcd}\exp(-l\cdot x)\cos(m\cdot x)\nonumber\\
&&+V_{abcd}\exp(-l\cdot x)\sin(m\cdot x).\label{GWS1}
\ee
The appearance of $\exp(-l\cdot x)$ is due to the difference between Minkowskian and Euclidean signatures. 

Case 2: For $k\neq 0$ and $k\cdot k\neq0$, the solution space is 1-dimensional and $R_{abcd}$ converges to zero. 

Case 3: For $k=0$ the vector $\O(a)=(1,1,1,1)$ is a constant and $R_{abcd}$ converges to a nonzero constant. The solution space corresponds to the full 10-dimensional space of traceless tensors $W_{abcd}$.

The geometries in Case 1 are smooth solutions of linearized Einstein equation, as Euclidean analog of plane waves. They correspond to the nontrivial low energy excitations from SFM under SCL. Case 2 with $R_{abcd}=0$ doesn't change the flat background geometry and, thus, correspond to purely gauge fluctuations of the triangulation in the flat geometry.

The solutions in Case 3 deserves some further explanation. Although those solutions appear in addition to the ``plane wave'' geometries Eq.\Ref{GWS1}, they only associate to $k=0$. So the set of solutions in case 3 is of measure-zero in the space of all solutions. The space of all solutions in the continuum limit is infinite-dimensional, although the solution space with a fixed $k$ is finite-dimensional. A generic linear combination
\be
R_{abcd}(x)=\int_{\C^4} \rmd k\, \delta^4(k\cdot k)\,\mathrm{Re}\lt[W_{abcd}(k) \exp\lt(-k\cdot x\rt)\rt]
\ee   
is insensitive to the value of $W_{abcd}(0)$ (solution in Case 3). The above $R_{abcd}(x)$ is a Euclidean analog of a realistic gravitational wave that is not a purely plane wave but has a distribution $W_{abcd}(k)$.

Among the zero modes mentioned in Section \ref{EOLT}, 4 diffeomorphisms have been taken care in the above analysis because of using deficit angle variables, which leads to $\pm2$ helicities. The hyperdiagonal zero mode has the same behavior as in Case 2, i.e. it converges to zero curvature $R_{abcd}=0$ \cite{0264-9381-5-12-007}.





\section{Some Topological Properties of the Triangulation}\label{TPOT}

\begin{figure}[h]
	\centering\includegraphics[width=0.6\linewidth]{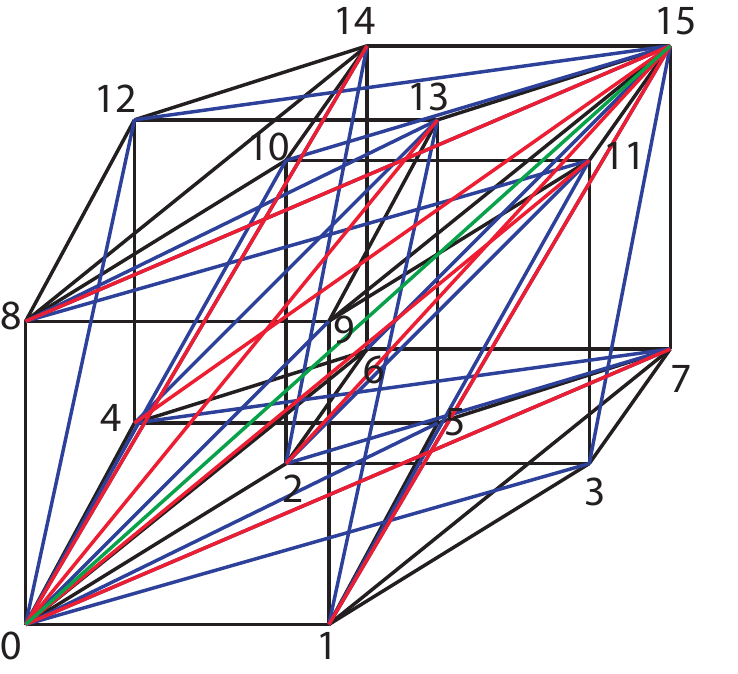}
	\caption{A visualization of a triangulated hypercube cell.The vertices of the hypercube are labeled by number from $0$ to $15$. The binary number of the vertex label is the same as the components of the vector from the origin point to the vertex.
	}\label{fig:Hypercube}
\end{figure}

The analysis in this paper is based on a fixed type of triangulation $\ck$. In this section we collect a couple of useful properties of $\ck$.  

$\ck$ is adapted to a 4-dimensional hypercube lattice in which each lattice cell is a triangulated hypercube (FIG.\ref{fig:Hypercube}). Each vertex of the hypercube is labelled by a number from $0$ to $15$. Note that the vertex number written in binary form $(n_1,n_2,n_3,n_4)$ with $n_i=0,1$ yields the components of the vector from the origin to the vertex. Thus the vertex numbers define 15 lattice vectors  at the origin, which are edges and various diagonals of the hypercube and subdivide the hypercube into 24 4-simplices. 
The triangulation $\ck$ is made from the hypercube lattice by simply translating the triangulation from one hypercube to another. In order to simplify the problem, one can consider $\ck$ as a $N^4$ lattice. Among those hypercubic cells, a hypercube whose lattice components contain $0$ or $N-1$ lies on the boundary of the lattice.  A hypercube whose lattice components  do not contain $0$ or $N-1$ is in the bulk.


A single triangulated hypercube has 65 edges, 110 triangles and 24 4-simplices. However, the numbers of the edges and the triangles per bulk cell in the lattice is smaller than those numbers for a single hypercube since triangles and edges are shared by different hypercube cells.
If there are $n$ edges or triangles parallels to each other in a single triangulated hypercube then each of those edges or triangles will be shared by $n$ hypercube cells in the bulk of the lattice.  Thus the  effective weight of those edges or faces in a cell is $1/n$. 

For example, in a single hypercube  the triangle $(4,5,15)$ is the only triangle that is parallel to $(0,1,11)$. One finds that the shift vector between $(0,1,11)$ and $(4,5,15)$ is $(0,1,0,0)$. In the bulk of the lattice, the triangle $(4,5,15)$ of cell with lattice coordinate $(t,x-1,y,z)$ coincides with the triangle $(0,1,11)$ of the cell $(t,x,y,z)$. Similarly, the triangle $(0,1,11)$ in the cell $(t,x+1,y,z)$ coincides with $(4,5,15)$ in cell $(t,x,y,z)$. Thus the bulk cell $(t,x,y,z)$ only posses half of the triangle $(0,1,11)$ and half of $(4,5,15)$.
Similar arguments work for all the other faces and edges in the bulk of the lattice $\ck$.
So in the lattice, each bulk hypercube only posses $15$ edges and $50$ triangles. 

Furthermore, we can define a coincide number $\psi$ of a triangle $f$ where  $\psi=m+1$ if one triangle $f$ coincide with $m$ triangles coming from other cells.  The maximum value of $\psi(f)$ is equal to one plus the number of the triangles that are parallel to $f$ in a single isolated hypercube\footnote{The maximum value of $\psi(f)$ also equals one over the weight of the triangle $f$.}. For any triangle $f$ in a bulk cell, $\psi(f)$ must be equal to its maximum value. But in a boundary cell, not all the triangles have maximum $\psi(f)$. Those triangles lie in the boundary triangles.

In an $N^4$ lattice, the boundary hypercubes contribute $356+574(N-2)+310(N-2)^2+56(N-2)^3$ boundary triangles and $80+148(N-2)+84(N-2)^2+14(N-2)^3$ boundary edges. So in the bulk, there are $50N^4-(356+574(N-2)+310(N-2)^2+56(N-2)^3)$ triangles and $15N^4-(80+148(N-2)+84(N-2)^2+14(N-2)^3)$ edges. When $N$ tends to be large, the ratio between the number of bulk edges and the number of bulk triangles will converge to $3:10$ .

Furthermore one can show that every bulk triangle is shared by an even number of 4-simplices because any triangle within a single  triangulated hypercube must be shared by 1,2,4 or 6 4-simplices. 
Define $\tilde{n}(f)$ to be  the total number of 4-simplices within a hypercube that are sharing the triangle $f$. We call $f$ of type-1 if $\tilde{n}(f)=1$, or of type-2 if $\tilde{n}(f)\neq1$ respectively. There are 24 type-1 triangles in a single hypercube. TABLE.\ref{table:1} lists all of those triangles and the triangles parallel to them.
\begin{widetext}
	\begin{center}
		\begin{table}[htbp]
			\caption{Each column of the table shows 4 triangles that parallel to each other. The triangles appears in the first two lines are type-1 and the triangles in the last two lines are type-2. }\label{table:1}
			\begin{tabular}{|c|c|c|c|c|c|c|c|c|c|c|c|c|}
				\hline
				type-1&(1,5,13)&(1,3,7)&(1,3,11)&(1,9,13)&(1,9,11)&(2,6,7)&(2,3,7)&(2,3,11)&(2,10,11)&(4,5,7)&(4,6,7)\\
				\hline
				type-1&(2,6,14)&(8,10,14)&(4,6,14)&(2,10,14)&(4,12,14)&(8,12,13)&(8,9,13)&(4,5,13)&(4,12,13)&(8,9,11)&(8,10,11)\\
				\hline
				type-2&(0,4,12)&(0,2,6)&(5,7,15)&(3,11,15)&(5,13,15)&(10,14,15)&(10,11,15)&(0,1,9)&(0,8,9)&(0,1,3)&(0,2,3)\\
				\hline
				type-2&(3,7,15)&(9,11,15)&(0,2,10)&(0,8,12)&(0,8,10)&(0,4,5)&(0,1,5)&(6,7,15)&(6,14,15)&(12,13,15)&(12,14,15)\\
				\hline
			\end{tabular}
		\end{table}
	\end{center}
\end{widetext}

Obviously some of the triangles are shared by different hypercubes. For those triangles one should add up $\tilde{n}(f)$ in different hypercubes in order to count how many 4-simplices are sharing the face $f$. TABLE.\ref{table:1} shows that each of the type-1 triangle must be parallel to another type-1 triangle and two type-2 triangles. From this we may conclude: 
\begin{itemize}
	\item Any triangle shown in the TABLE.\ref{table:1} is shared by 4 hypercubes. In two of those hypercubes, the triangle  is type-1 and in the other two hypercubes, it is type-2.
	\item The triangles listed in the same column are shared by the same number of 4-simplices. Explicitly, the triangle $(x,y,z)$ is shared by $\sum_f \tilde{n}(f)$ of 4-simplices, where $f$ stands for all the triangles that are in the column and contain triangle $(x,y,z)$. Moreover, $\sum_f \tilde{n}(f)$ must be even since it can be expressed as $1+1$ plus two even number.
	\item For the other type-2 triangle in the TABLE.\ref{table:1}, the number of 4-simplices shared by it should be the sum of $2,4$ or $6$, which is also even. 
\end{itemize}

Thus in the bulk of $\ck$, every triangle is shared by an even number of 4-simplices.


\end{document}